\documentclass[useAMS,usenatbib]{mn2e}
\usepackage{amsmath}
\usepackage{amssymb}
\usepackage{multirow}
\usepackage{graphicx}
\usepackage{epstopdf}
\usepackage{cancel}

\graphicspath{ {figs/} }

\def\simlt{\lower.5ex\hbox{$\; \buildrel < \over \sim \;$}}
\def\simgt{\lower.5ex\hbox{$\; \buildrel > \over \sim \;$}}
\newcommand{\Msun}{M_\odot}
\newcommand{\SurveyRedshift}{1.35}
\newcommand{\SurveyRedshiftLowz}{1.25}
\newcommand{\SurveyRedshiftHighz}{1.58}

\newcommand{\SurveyDelayMin}{1.5}
\newcommand{\SurveyDelayMax}{3.7}
\newcommand{\SurveyDelayLowzMin}{2.4}
\newcommand{\SurveyDelayLowzMax}{3.8}
\newcommand{\SurveyDelayHighzMin}{1.5}
\newcommand{\SurveyDelayHighzMax}{2.7}
\newcommand{\TotNumofSNe}{29}
\newcommand{\TotSNeRateCount}{13}
\newcommand{\TotSNeRateCountSysErr}{2}
\newcommand{\TotSNeRateCountStatErr}{3.6}
\newcommand{\TotSNeRateCountHighz}{5.5}
\newcommand{\TotSNeRateCountHighzSysErr}{3.8}
\newcommand{\TotSNeRateCountLowz}{7.5}
\newcommand{\TotSNeRateCountLowzSysErr}{3.2}
\newcommand{\NumofSurveyIas}{11}
\newcommand{\NumofSurveyPossibleIas}{4}
\newcommand{\XavierAlpha}{0.156}
\newcommand{\XavierBeta}{2.460}
\newcommand{\OldHighz}{1.1}
\newcommand{\OldShortDelay}{3.2}
\newcommand{\SurveyIaRate}{2.6}
\newcommand{\SurveyIaRateUpErr}{3.2}
\newcommand{\SurveyIaRateDnErr}{1.5}
\newcommand{\SurveyIaRateLowz}{2.2}
\newcommand{\SurveyIaRateHighz}{3.5}
\newcommand{\SurveyIaRateLowzUpErr}{2.6}
\newcommand{\SurveyIaRateLowzDnErr}{1.3}
\newcommand{\SurveyIaRateHighzUpErr}{6.6}
\newcommand{\SurveyIaRateHighzDnErr}{2.8}
\newcommand{\SurveyAlpha}{-1.30}
\newcommand{\SurveyAlphaUErr}{0.23}
\newcommand{\SurveyAlphaLErr}{0.16}
\newcommand{\AlphaTwoParamOneBin}{-1.45}
\newcommand{\AlphaTwoParamOneBinUErr}{0.51}
\newcommand{\AlphaTwoParamOneBinLErr}{0.38}
\newcommand{\AlphaOneParamOneBin}{-1.30}
\newcommand{\AlphaOneParamOneBinUErr}{0.23}
\newcommand{\AlphaOneParamOneBinLErr}{0.16}
\newcommand{\AlphaOneParamTwoBin}{-1.30}
\newcommand{\AlphaOneParamTwoBinUErr}{0.23}
\newcommand{\AlphaOneParamTwoBinLErr}{0.16}
\newcommand{\AlphaTwoParamTwoBin}{-1.42}
\newcommand{\AlphaTwoParamTwoBinUErr}{0.46}
\newcommand{\AlphaTwoParamTwoBinLErr}{0.34}
\newcommand{\NMsOneBin}{9.0}
\newcommand{\NMsOneBinUErr}{29.2}
\newcommand{\NMsOneBinLErr}{7.1}
\newcommand{\NMsTwoBin}{8.1}
\newcommand{\NMsTwoBinUErr}{21.0}
\newcommand{\NMsTwoBinLErr}{6.0}

\title[Cluster SN rate and DTD to $z=1.75$]{The rate of Type-Ia supernovae in galaxy clusters and the delay-time distribution out to redshift 1.75}
\author[M. Friedmann and D. Maoz]{
Matan Friedmann\thanks{E-mail: matanfri@mail.tau.ac.il (MF)}
and Dan Maoz
\\
School of Physics and Astronomy, Tel Aviv University, Tel Aviv 69978, Israel
}

\date{Accepted XXX. Received YYY; in original form ZZZ}

\begin{document}
\pagerange{\pageref{firstpage}--\pageref{lastpage}} \pubyear{2018}
\maketitle

\label{firstpage}

\begin{abstract}
  The observed delay-time distribution (DTD) of Type-Ia supernovae (SNe Ia) is a valuable probe of SN Ia progenitors and physics, and of the role of SNe Ia in cosmic metal enrichment. 
  The SN Ia rate in galaxy clusters as a function of cluster redshift is an almost-direct measure of the DTD, but current estimates have been limited out to a mean redshift $\langle z \rangle=\OldHighz$, corresponding to time delays, after cluster star-formation, of $\gtrsim \OldShortDelay$~Gyr. 
  We analyze data from a Hubble Space Telescope monitoring project of 12 galaxy clusters at $z=1.13-1.75$, where we discover \TotNumofSNe ~SNe, and present their multi-band light curves. 
  Based on the SN photometry and the apparent host galaxies, we assess cluster membership and SN type, finding \NumofSurveyIas ~cases that are likely SNe~Ia in cluster galaxies and \NumofSurveyPossibleIas ~more cases which are possible but not certain cluster SNe~Ia. 
We conduct simulations to estimate the SN detection efficiency, the experiment's completeness, and the photometric errors, and perform photometry of the cluster galaxies to derive the cluster stellar masses. 
  With this input, we obtain a mean $\langle z \rangle=\SurveyRedshift$ cluster rest-frame SN~Ia rate per unit formed stellar mass of $R_{\rm Ia, m*}=\SurveyIaRate ^{+ \SurveyIaRateUpErr} _{- \SurveyIaRateDnErr} \times 10^{-13} {\rm ~yr}^{-1} {\rm M}_\odot^{-1}$. 
  Separating the cluster sample into high-$z$ and low-$z$ bins, the rates are $\SurveyIaRateLowz ^{+ \SurveyIaRateLowzUpErr} _{- \SurveyIaRateLowzDnErr} \times 10^{-13} {\rm ~yr}^{-1} {\rm M}_\odot^{-1} $ at $\langle z \rangle=\SurveyRedshiftLowz$, and $\SurveyIaRateHighz ^{+ \SurveyIaRateHighzUpErr} _{- \SurveyIaRateHighzDnErr} \times 10^{-13} {\rm ~yr}^{-1} {\rm M}_\odot^{-1}$ at $\langle z \rangle=\SurveyRedshiftHighz$.
  Combining our results with previous cluster SN~Ia rates, we fit the DTD, now down to delays of 1.5~Gyr, with a power-law dependence, $t^\alpha$, with $\alpha=\SurveyAlpha ^{+\SurveyAlphaUErr} _{-\SurveyAlphaLErr}$. We confirm previous indications for a SN~Ia production efficiency that is several times higher in galaxy clusters than in the field, perhaps caused by a peculiar stellar initial mass function in clusters, or by a higher incidence of binaries that will evolve into SNe~Ia.
\end{abstract}


\section{Introduction}
\label{Sec:Intro}

Type-Ia supernovae (SNe~Ia) play numerous important roles in astrophysics, including as the distance indicators that first revealed the accelerating universal expansion \citep{Riess1998, Perlmutter1999}, and continue to serve as a precision cosmological probe (e.g. \citealt{Betoule2014,Scolnic2018}). 
Disturbingly, however, major questions remain regarding the identity of the progenitor systems that eventually explode as SNe~Ia (see \citealt{Maoz2014,Livio2018}, for recent reviews) and regarding the physics of the explosions themselves (see \citealt{Hillebrandt2013}). 
The so-called delay-time distribution (DTD) of SNe~Ia has emerged as a powerful discriminant among different models. 
The DTD is the hypothetical SN~Ia rate, as a function of time, that would follow a short burst of star formation that formed a unit stellar mass. 
Among the various observational approaches that have been used to recover the DTD (see \citealt{MaozMannucci2012,Maoz2014}), an almost-direct approach is to measure the SN~Ia rate in massive clusters of galaxies at diverse redshifts \citep{Maoz2010}. 
Observations of the stellar populations in cluster galaxies show that the bulk of their stars were formed in brief episodes at redshifts between $z=3-4$ (e.g., \citealt{Daddi2000,Stanford2005,Eisenhardt2008,Snyder2012,Andreon2014}, and \citealt{Stalder2013,Andreon2016} for some of the very same clusters analysed here; see \citealt{Miller2018,Oteo2018} for recent examples of clusters in the process of forming stars at those epochs.)
A measurement of the SN~Ia rate per formed stellar mass in a sample of galaxy clusters in some small redshift interval therefore gives an estimate of the DTD at a delay corresponding to the time elapsed between the redshift of star formation and the observed redshift of the sample. 

\citet{Maoz2017} recently re-compiled and compared the various existing SN~Ia DTD measurements for both field and cluster galaxies. 
In field galaxies, the best estimate for the DTD's functional form is a power-law, $t^{-1.1\pm 0.1}$. 
In clusters, the DTD is consistent with the same form but is less well-constrained, $t^{-1.3\pm 0.4}$, because cluster SN rates have been measured no further back in cosmic time than mean redshift $\langle z \rangle\sim \OldHighz $ \citep{Barbary2012}, corresponding to delay times $t\gtrsim \OldShortDelay$~Gyr after star formation.

In terms of the normalisation of the DTD, for field galaxies the Hubble-time-integrated (between delays of 40~Myr and 13.7~Gyr) number of SNe~Ia per unit formed stellar mass is $N_{\rm Ia}/M_\star=1.3\pm 0.1\times 10^{-3} {\rm M}_\odot^{-1}$ (as in \citet{Maoz2017}, here and forthwith all stellar mass estimates are under the assumption of a \citet{Kroupa2001} initial mass function (IMF)).
\citet{Maoz2017} re-affirmed that cluster galaxies, based on their high DTD values at the measured delays of $> \OldShortDelay$~Gyr, appear to have a production efficiency of SNe that is several times higher than that of field galaxies at the same delays.
This result might be related to the long-known puzzle of the large iron-to-stellar mass ratio seen in galaxy clusters, observed by means of X-ray emission of the hot intracluster gas (e.g. \citealt{Maoz2010, Loewenstein2013, RenziniAndreon2014}). 
Considering the measured cluster SN~Ia rates and the iron constraints, \citet{Maoz2017} estimate, for galaxy clusters, $N_{\rm Ia}/M_\star=5.4\pm 2.3\times 10^{-3}~{\rm M}_\odot^{-1}$. 
Finally, \citet{Maoz2017} found additional circumstantial evidence for the reality of an efficient DTD that is unique to cluster-like environments. 
They showed that such a DTD, operating in conjunction with a brief burst of star-formation at $z\sim 3$ (i.e. similar to the star-formation history of galaxy clusters) can reproduce the pattern of abundance ratios of $\alpha$-elements, iron, and hydrogen, seen in the ``high-$\alpha$'' locus of halo stars in the Milky Way, a population that could be in some way similar to the stellar populations in galaxy clusters.

From a physical perspective, a high-normalised SN~Ia DTD in galaxy-cluster-like environments could arise in various ways. 
For example, star formation in such environments could proceed via a ``middle-heavy'' (or ``paunchy'') IMF (e.g. \citealt{Fardal2007}) with an excess of $\sim 3-8~{\rm M}_{\odot}$ stars, relative to $1~{\rm M}_{\odot}$ stars (the stars that dominate the light in old stellar populations, and hence are the ones that are actually observed), hence leading to a larger number of white dwarfs (white dwarfs are the fundamental ``explosive'' in all SN~Ia models).
Alternatively, star formation and binary evolution in cluster environments could
perhaps lead to a larger number of close-binary pairs that eventually explode as SNe~Ia \citep{MaozHallakoun2018}.
However, some possibility remains that the high SN~Ia rates measured to date in galaxy-cluster SN surveys have been over-estimated because of some systematic error. 
Furthermore, even if the high rates are real, the extrapolation of the DTD back to short delays is not necessarily correct, and some other source may be behind the large mass of iron in the intracluster medium (e.g. some population of high-iron-yield core-collapse SNe that is peculiar to such environments). 
To test these possibilities, it is important to extend measurements of the DTD in clusters to shorter delays, by measuring the SN~Ia rate out to higher redshifts, closer to the time of star formation in clusters. 
Such measurements, if they reach early enough in cosmic time, can potentially even reveal ``in action'' the players actually responsible for the bulk of iron enrichment, be they SNe~Ia, or normal or peculiar core-collapse SNe.

Recent surveys have led to the discovery of galaxy clusters in the redshift range $z\sim 1-2$ (e.g. \citealt{Reichardt2013,Stanford2014}).
The identification of these high-$z$ clusters has prompted observations with the {\it Hubble Space Telescope} (HST) to search for SNe~Ia (e.g. \citealt{Rubin2017}) for cosmological applications (e.g. constraining the cosmic equation of state) and for gravitational lensing studies of the clusters \citep{Jee2017}. 
In this paper, we analyze the archival HST data from a high-$z$ cluster monitoring program, to derive the cluster SN~Ia rate out to redshift $z=1.75$. 
We then combine our SN~Ia rate measurement with previously measured cluster rates to obtain new constraints and insight on the SN~Ia DTD and on its physical implications.

\section{Cluster Sample and Observations}
\label{Sec:Obs}
Our analysis is based on archival data from a multi-epoch HST imaging program of a sample of 12 massive galaxy clusters at redshifts between 1.13 and 1.75 (HST Programs GO-13677 and GO-14327, PI: S. Perlmutter, Observing Cycles 22 and 23).
Table \ref{tab:cluster_sample_table} lists basic parameters for the 12 clusters, the dates of the HST individual observing season or seasons for each cluster, and references to the cluster discovery and characterisation papers.

\begin{table*}
	\centering
	\caption{Galaxy cluster sample.}
	\label{tab:cluster_sample_table}
	\begin{tabular}{cccccccc}
		\hline
		 Cluster ID & Nickname & Redshift & $\alpha$ (J2000) & $\delta$ (J2000) & Obs. Start & Obs. End &Ref. \\ 
		 (1) & (2) & (3) & (4) & (5) & (6) & (7) & (8) \\ 
		\hline
		 IDCS J1426.5+3508 &   IDCS1426 & 1.75 & $14:26:33$ & $+35:08:24$ & 2014/11/17 & 2015/04/10 & B12 \\ 
		& & & & & 2015/11/27 & 2016/07/28 &  \\ 
		 ISCS J1432.3+3253 &   ISCS1432 & 1.40 & $14:32:18$ & $+32:53:08$ & 2014/11/10 & 2015/08/07 & B13 \\ 
		& & & & & 2016/07/11 & 2017/03/30 &  \\ 
		    MOO J1014+0038 &    MOO1014 & 1.27 & $10:14:08$ & $+00:38:22$ & 2014/10/22 & 2015/06/01 & B15 \\ 
		& & & & & 2015/10/19 & 2016/05/05 &  \\ 
		    MOO J1142+1527 &    MOO1142 & 1.19 & $11:42:46$ & $+15:27:14$ & 2015/11/02 & 2016/05/15 & G15 \\ 
		 SpARCS J0224-0323 & SPARCS0224 & 1.63 & $02:24:28$ & $-03:23:32$ & 2015/06/13 & 2015/12/06 & N16 \\ 
		 SpARCS J0330-2843 & SPARCS0330 & 1.63 & $03:30:54$ & $-28:43:10$ & 2015/05/25 & 2016/03/16 & N16 \\ 
		 SpARCS J1049+5640 & SPARCS1049 & 1.70 & $10:49:23$ & $+56:40:33$ & 2014/09/15 & 2015/03/31 & W15 \\ 
		 SpARCS J0035-4312 & SPARCS3550 & 1.34 & $00:35:50$ & $-43:12:24$ & 2015/06/06 & 2015/11/27 & W09 \\ 
		& & & & & 2016/04/09 & 2016/07/23 &  \\ 
		 SPT-CL J0205-5829 &    SPT0205 & 1.32 & $02:05:46$ & $-58:29:07$ & 2014/10/06 & 2016/09/24 & S13 \\ 
		 SPT-CL J2040-4451 &    SPT2040 & 1.48 & $20:40:60$ & $-44:51:37$ & 2015/03/07 & 2015/12/02 & B14 \\ 
		& & & & & 2016/07/26 & 2016/07/26 &  \\ 
		 SPT-CL J2106-5844 &    SPT2106 & 1.13 & $21:06:05$ & $-58:44:42$ & 2015/03/07 & 2015/09/15 & F11 \\ 
		& & & & & 2016/02/28 & 2016/07/29 &  \\ 
		 XMMU J0044.0-2033 &      XMM44 & 1.58 & $00:44:05$ & $-20:33:44$ & 2015/05/09 & 2015/12/06 & S11 \\ 
		\hline
	\end{tabular}
\\
\raggedright{Notes---a second continuous observing season for a cluster is listed as a second row. References: B12--\citet{Brodwin2012}; B13--\citet{Brodwin2013}; B15--\citet{Brodwin2015};G15--\citet{Gonzalez2015}; N16--\citet{Nantais2016}; W15--\citet{Webb2015}; W09--\citet{Wilson2009}; S13--\citet{Stalder2013}; B14--\citet{Bayliss2014}; F11--\citet{Foley2011}; S11--\citet{Santos2011}.}
\end{table*}

\begin{figure}
	\includegraphics[width=\columnwidth]{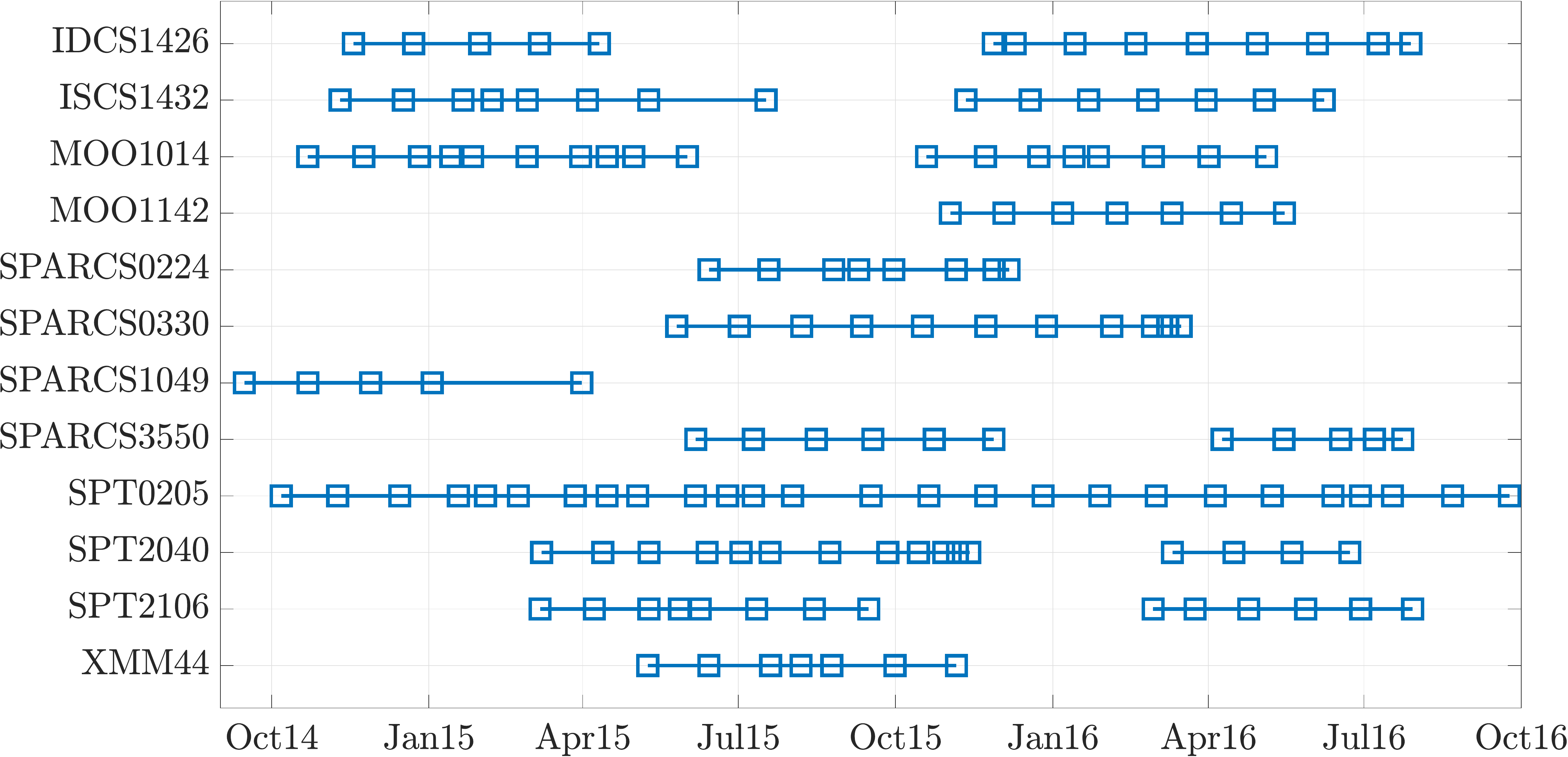}
        \caption{HST visits to the sample clusters during 2014-2016. Each square represents an epoch in the F140W band (except for cluster SPARCS1049, for which its F160W epochs are shown).}
    \label{fig:survey_cov}
\end{figure}

A total of 498 images were obtained, over 174 orbits, with the HST Wide Field Camera 3 (WFC3), using both its infrared channel (WFC3-IR, field of view $123''\times 136''$, pixel scale $0\farcs13$~pixel$^{-1}$), and the ultraviolet-visible channel (WFC3-UVIS, field of view $162''\times 162''$, pixel scale $0\farcs04$~pixel$^{-1}$). 
Filters used were F814W (with UVIS), and F105W, F125W, F140W and F160W (in IR), with pivot wavelengths of $0.80 ~\mu$m, $1.05~ \mu$m, $1.25~ \mu$m, $1.39~ \mu$m, and $1.54 ~\mu$m, respectively.
At the redshifts of the clusters, these bands correspond roughly to rest-frame optical bands from $U$ to $R$. 
Each of the WFC3-IR visits to a cluster target consisted of three individual dithered sub-exposures. 
The WFC3-UVIS F814W visits consisted of a single exposure.
HST observations were carried out between 2014, September 15th, and 2016, September 24th, with one additional observation on 2017, March 30th.
In a typical HST visit to a cluster field, F814W, F105W, and F140W exposures were all obtained. 
Exposures in F160W (and in one case F125W) were triggered at particular epochs for specific clusters by the program's proposers in Target-of-Opportunity mode, in reaction to SN candidates they detected in the routine survey images. 
The typical cadence between repeated visits to each field was about 30 observer-frame days. 
Figure~\ref{fig:survey_cov} shows the temporal coverage of the sample. 

\begin{figure*}
	\includegraphics[width=\textwidth]{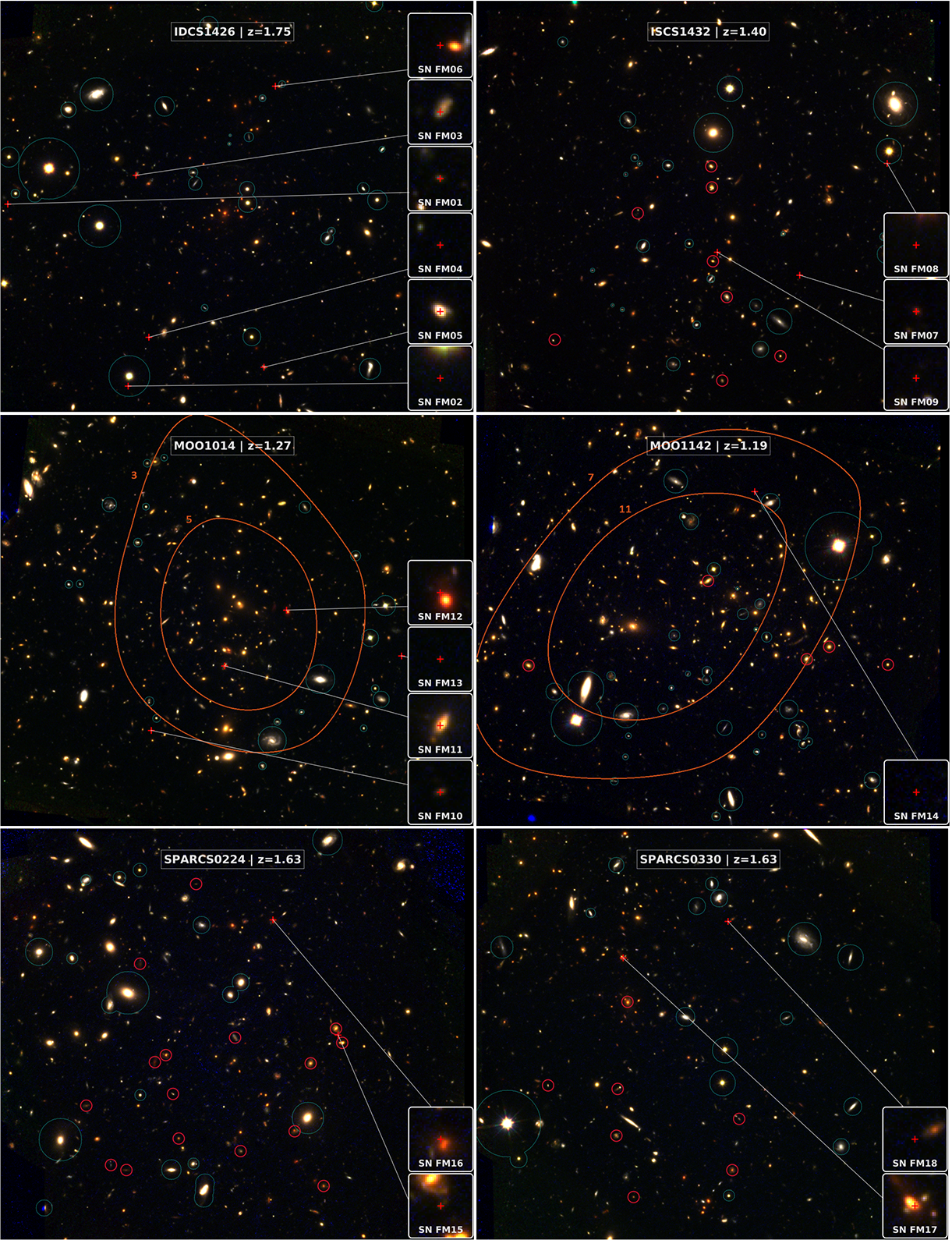}
        \caption{Color renditions of the galaxy clusters in the survey, produced by combining F814W, F105W, and F140W images.
Selected SZE contours from the cluster discovery papers, when available, are superimposed on the images, with the SZE signal-to-noise level marked along the contour. Sub-frames zoom in on the locations, marked with crosses, of detected candidate SNe. Red circles mark spectroscopically confirmed cluster members, and cyan circles indicate bright, likely foreground, objects.} 
    \label{fig:cluster_superplots}
\end{figure*}

Raw images were processed by the standard HST data pipeline, and the dithered sub-exposures of a visit were combined by the pipeline into a
final "drizzled" image with the MultiDrizzle algorithm \citep{MultiDrizzleBook2009}. 
Each drizzled image has sub-pixel resolution, and an effective exposure time of about 1000~s. Figure~\ref{fig:cluster_superplots} shows combined multi-band (F814W, F105W, and F140W) color renditions of each of the 12 clusters. To help visualize a cluster's center and extent, we overplot (when available in the literature) several selected Sunyaev-Zeldovich effect (SZE) contours.

\begin{figure*}
	\includegraphics[width=\textwidth]{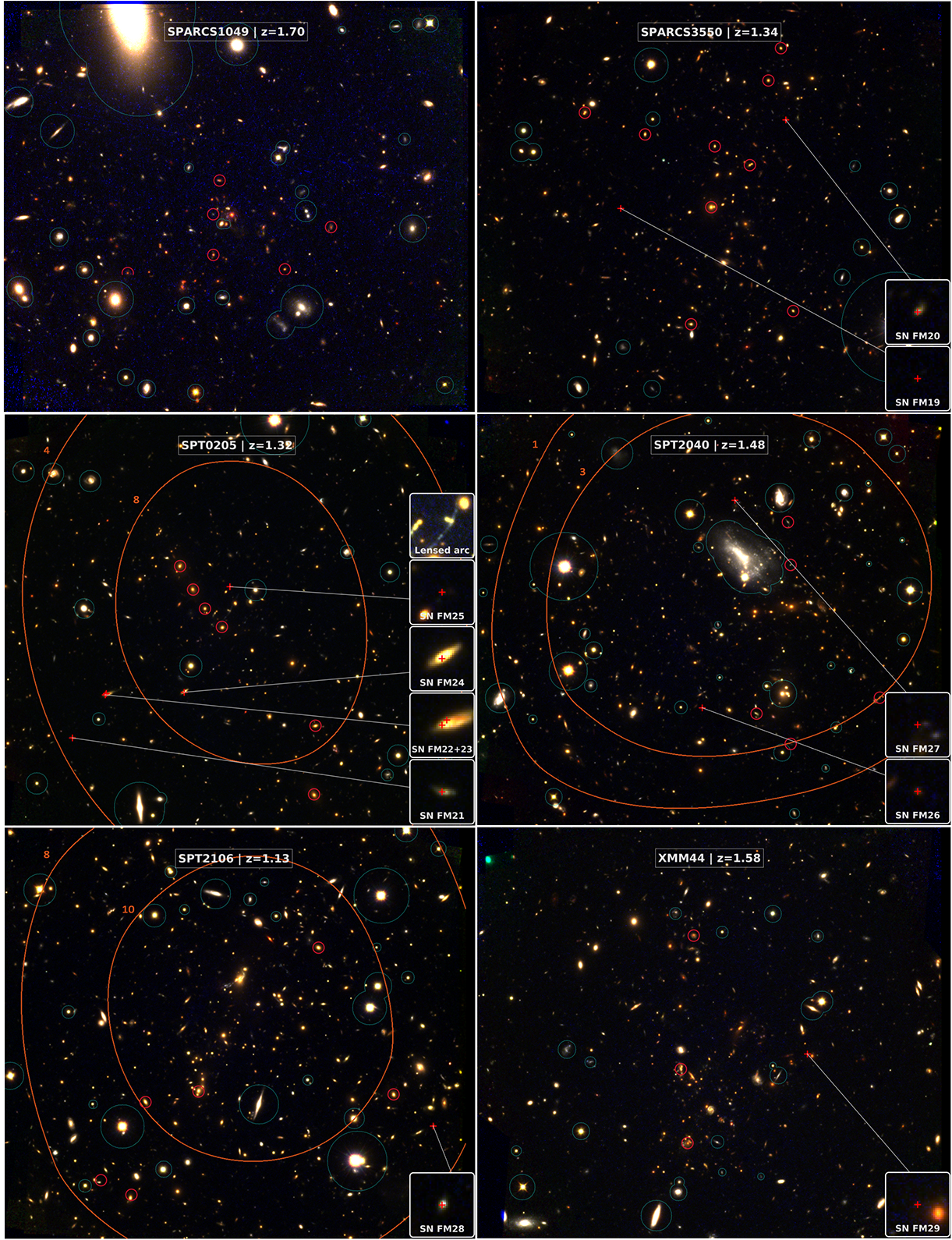}
        \caption{Fig. \ref{fig:cluster_superplots}, continued.}
    \label{fig:cluster_superplots2}
\end{figure*}

\section{Supernova detection and photometry}
\label{Sec:sn_det}
For the purpose of searching for transients in the HST data, all images of every cluster field were rotated (based on the telescope orientation as listed in the image headers) and shifted (based on the results of a cross correlation between images) to register them with the first-epoch image of that cluster field. 
A deep reference image was obtained by forming the mean of all the coaligned images in each filter for every field, while applying sigma-clipping rejection of the pixels in every pixel stack that are dead, hot, or affected by cosmic-ray hits. 
Difference images were formed for each epoch in every filter by subtracting the appropriate reference image. 
Owing to the stable HST point-spread function (PSF), no PSF matching between images being subtracted was required in order to obtain satisfactory difference images.
The diverse image orientations at different epochs resulted in a central region of each cluster field that had full time coverage (i.e. no missing epochs) and the greatest image depth, surrounded by shallower areas, having partial time coverage, along the field edges. 
Although we have discovered SN candidates throughout the field, including the edge regions, and we report also those candidates, to simplify our efficiency and completeness calibrations we consider only the transients and the galaxies within the central full-coverage area for our actual SN rate measurement calculations. 

Among the various observed bands, we decided to perform our search for transients in the F140W band since it turned out, after some experimentation, that all candidate SNe that we found are always detected in this band, and at a higher S/N ratio than in the other IR band used in the routine survey, F105W. 
For the SPARCS1049 cluster field, which was not observed by the program in F140W, we performed the transient search using its F160W images. 
Transient candidates were identified by searching for point-source-like residuals in the difference images of each epoch. 
The search consisted of an automated first screening, followed by a visual inspection by one of us (MF). 
For the first step, in each difference image the artificial residuals associated with bright foreground objects (stars and bright nearby galaxies) were automatically masked out. 
The images were then smoothed with a running-mean box-car filter of size of $3\times 3$ pixels, which further reduced the effect of artifacts. 
Finally, all pixels with counts $5\sigma$ above the local background fluctuations were clustered by means of a hierarchical clustering algorithm, resulting in a set of initial candidates from this automated detection phase (with typically around 50-90 such automated detections per epoch of a field).
All automated candidate detections were then classified by eye, either as noise or as possible real transients, using a graphic user interface (GUI) built for this purpose. 
To avoid any subjective bias, the decision on a candidate's reality was independent for each epoch, and only after all detections were vetted were we able to realize whether or not several detections at different epochs were, in fact, at the same location in the field. 
To carefully investigate each possible detection, the GUI included various image visualisation and manipulation tools, e.g. the ability for fine realignment of a small image section, to obtain an improved difference image at the location of the transient candidate. 

The reference images used for the image subtraction were composed of all epochs, i.e. also epochs that included signal from the detected transient sources.
Before performing photometry on the transients, and in order to avoid the subtraction of some of the transient source's light itself, we created new reference images for every candidate transient, with each image stack including only epochs in which the candidate
was undetected. 
For every transient source detected in the F140W search, we searched the same position in the F814W and F105W difference images, and in any other existing bands at all epochs (e.g. Target-of-Opportunity F160W exposures obtained for some epochs).
We performed photometry on the difference images of each detected transient by means of PSF fitting. 
The HST PSF for each band was simulated using \textsc{Tiny Tim}\footnote{http://www.stsci.edu/hst/observatory/focus/TinyTim} (\citealt{TinyTimRef}). 
The PSF was scaled to fit by $\chi^2$ minimisation both the detected transient point source, and a bright (but unsaturated) star in a deep reference image. 
The ratio of the PSF scalings times the total count rate within a large aperture around the bright star gave the transient's count rate.
We used the PHOTFLAM and PHOTFNU keywords in the HST image headers to
convert count rates to physical units of flux density within a given band, $f_{\lambda}~[{\rm erg} \ {\rm cm}^{-2}~ {\rm s}^{-1}{\rm \text{\AA}}^{-1}]$, $f_{\nu}~[{\rm erg} \ {\rm cm}^{-2}~ {\rm s}^{-1}{\rm Hz}^{-1}]$, and AB magnitudes. 
The photometric errors were estimated by means of simulated point sources that we planted in the real images, recovered through our detection pipeline, and then photometered, as further described in Section~\ref{Sec:det_eff}, below. 

In three cases of images obtained through the "followup" filters, we did not have SN-free epochs at the location of a transient, and therefore could not create a reference image for subtraction. 
In these cases, we instead subtracted the reference image in the nearest available band that had clean observations of the host, after scaling it to match the brightness of the host in the band in question. Subtraction of the reference image then permitted PSF photometry, as above, of the residual transient point source.

\section{Candidate supernova sample and classification}
\label{Sec:sn_class}
Following the above procedures, we have identified \TotNumofSNe ~transients that we consider likely to be real SN detections, which we list in 
Table~\ref{tab:sne_table}.
 \begin{table*}
	\centering
	\caption{Supernova candidates. }
	\label{tab:sne_table}
	\begin{tabular}{ccccccccccc}
		\hline
ID & $\alpha$(J2000) & $\delta$(J2000) & Sep. & Dist. & Cl.Ia? & $x_1$ & $c$ & $M_B$ & $t_0$ & $\chi^2$(dof) \\ 
		  &  &  & [arcsec] & [kpc] &  &  &  & [mag] & (MJD) &  \\ 
		 (1) & (2) & (3) & (4) & (5) & (6) & (7) & (8) & (9) & (10) & (11) \\ 
		\hline
		\multicolumn{11}{|c|}{IDCS1426 [z=1.75]} \\
		\hline
		SN~FM01 & $14:26:38.3$ & $+35:08:26.6$ & 0.1 & 570 & $-$ & - & - & - & - & - \\ 
		SN~FM02 & $14:26:35.3$ & $+35:07:31.7$ & - & 450 & $-$ & - & - & - & - & - \\ 
		SN~FM03 & $14:26:35.2$ & $+35:08:35.0$ & 0.4 & 260 & $?$ & $  1.8 \pm  2.0$ & $ 0.21 \pm 0.12$ & $-19.16 ^{+0.16} _{-0.22}$ & $ 6975 \pm   10$ &    3.44(12) \\ 
		SN~FM04 & $14:26:34.8$ & $+35:07:46.5$ & 0.0 & 390 & $-$ & - & - & - & - & - \\ 
		SN~FM05 & $14:26:32.0$ & $+35:07:37.3$ & 0.1 & 420 & $-$ & - & - & - & - & - \\ 
		SN~FM06 & $14:26:31.7$ & $+35:09:02.0$ & 0.9 & 330 & $+$ & $ -1.3 \pm  1.9$ & $ 0.09 ^{+0.12} _{-0.15}$ & $-18.97 \pm 0.13$ & $ 6982 ^{+   6} _{-  10}$ &    8.38(11) \\ 
		\hline
		\multicolumn{11}{|c|}{ISCS1432 [z=1.40]} \\
		\hline
		SN~FM07 & $14:32:16.1$ & $+32:52:49.4$ & 0.1 & 290 & $+$ & $  0.4 \pm  1.5$ & $ 0.31 \pm 0.13$ & $-18.69 \pm 0.15$ & $ 7217 _{-   4}$ &    1.61(14) \\ 
		SN~FM08 & $14:32:13.9$ & $+32:53:23.1$ & - & 440 & $-$ & - & - & - & - & - \\ 
		SN~FM09 & $14:32:18.0$ & $+32:52:56.3$ & 0.1 & 120 & $-$ & - & - & - & - & - \\ 
		\hline
		\multicolumn{11}{|c|}{MOO1014 [z=1.27]} \\
		\hline
		SN~FM10 & $10:14:09.1$ & $+00:37:48.9$ & 0.1 & 340 & $+$ & $  0.1 \pm  0.9$ & $ 0.40 \pm 0.10$ & $-18.43 \pm 0.16$ & $ 6964 \pm    3$ &   22.54(19) \\ 
		SN~FM11 & $10:14:07.6$ & $+00:38:08.3$ & 0.1 & 110 & $+$ & $  2.1 \pm  0.8$ & $ 0.50 \pm 0.10$ & $-18.48 \pm 0.14$ & $ 7103 \pm    2$ &    9.88(18) \\ 
		SN~FM12 & $10:14:06.3$ & $+00:38:25.3$ & 0.6 & 140 & $-$ & - & - & - & - & - \\ 
		SN~FM13 & $10:14:04.0$ & $+00:38:11.4$ & - & 440 & $-$ & - & - & - & - & - \\ 
		\hline
		\multicolumn{11}{|c|}{MOO1142 [z=1.19]} \\
		\hline
		SN~FM14 & $11:42:44.9$ & $+15:27:52.1$ & - & 340 & $-$ & - & - & - & - & - \\ 
		\hline
		\multicolumn{11}{|c|}{SPARCS0224 [z=1.63]} \\
		\hline
		SN~FM15 & $02:24:26.2$ & $-03:23:32.2$ & 2.0 & 260 & $?$ & $ -1.4 \pm  0.4$ & $-0.23 \pm 0.03$ & $-19.74 \pm 0.04$ & $ 7265 \pm    2$ &   51.87(22) \\ 
		SN~FM16 & $02:24:27.6$ & $-03:22:57.4$ & 0.4 & 310 & $+$ & $ -2.7 \pm  1.3$ & $-0.03 ^{+0.15} _{-0.20}$ & $-19.03 ^{+0.22} _{-0.32}$ & $ 7324 \pm    6$ &    7.12(20) \\ 
		\hline
		\multicolumn{11}{|c|}{SPARCS0330 [z=1.63]} \\
		\hline
		SN~FM17 & $03:30:55.9$ & $-28:42:45.8$ & 0.3 & 290 & $?$ & $  4.0 _{- 0.5}$ & $ 0.41 \pm 0.03$ & $-19.00 \pm 0.09$ & $ 7428 ^{+   6} _{-   3}$ &   19.93(12) \\ 
		SN~FM18 & $03:30:53.5$ & $-28:42:34.8$ & 1.5 & 300 & $-$ & - & - & - & - & - \\ 
		\hline
		\multicolumn{11}{|c|}{SPARCS3550 [z=1.34]} \\
		\hline
		SN~FM19 & $00:35:52.1$ & $-43:12:24.0$ & - & 220 & $-$ & - & - & - & - & - \\ 
		SN~FM20 & $00:35:47.6$ & $-43:11:57.5$ & 0.2 & 300 & $?$ & $ -4.0 ^{+ 1.5}$ & $ 0.09 ^{+0.15} _{-0.03}$ & $-18.54 ^{+0.19} _{-0.10}$ & $ 7566 \pm    2$ &    2.05(10) \\ 
		\hline
		\multicolumn{11}{|c|}{SPT0205 [z=1.32]} \\
		\hline
		SN~FM21 & $02:05:52.3$ & $-58:29:40.0$ & 0.2 & 510 & $-$ & - & - & - & - & - \\ 
		SN~FM22 & $02:05:51.0$ & $-58:29:27.2$ & 1.0 & 370 & $+$ & $  1.9 \pm  0.4$ & $ 0.06 \pm 0.03$ & $-19.54 \pm 0.04$ & $ 7187 \pm    1$ &   39.15(29) \\ 
		SN~FM23 & $02:05:51.0$ & $-58:29:26.6$ & 0.6 & 370 & $+$ & $ -0.7 \pm  0.5$ & $-0.11 \pm 0.06$ & $-19.56 \pm 0.08$ & $ 7589 \pm    1$ &   14.80(15) \\ 
		SN~FM24 & $02:05:48.0$ & $-58:29:26.5$ & 0.3 & 220 & $+$ & $  1.4 \pm  0.7$ & $ 0.31 \pm 0.08$ & $-18.84 \pm 0.10$ & $ 7233 \pm    1$ &   38.79(27) \\ 
		SN~FM25 & $02:05:46.2$ & $-58:28:54.8$ & 0.3 &  90 & $+$ & $  1.5 \pm  0.4$ & $ 0.06 \pm 0.03$ & $-19.48 \pm 0.05$ & $ 7072 \pm    1$ &   83.18(31) \\ 
		\hline
		\multicolumn{11}{|c|}{SPT2040 [z=1.48]} \\
		\hline
		SN~FM26 & $20:40:59.7$ & $-44:52:04.1$ & 0.1 & 230 & $+$ & $  1.3 \pm  0.8$ & $ 0.13 \pm 0.08$ & $-19.29 \pm 0.09$ & $ 7301 \pm    2$ &   13.76(20) \\ 
		SN~FM27 & $20:40:58.8$ & $-44:51:01.6$ & 0.5 & 310 & $-$ & - & - & - & - & - \\ 
		\hline
		\multicolumn{11}{|c|}{SPT2106 [z=1.13]} \\
		\hline
		SN~FM28 & $21:05:57.1$ & $-58:45:11.9$ & 0.1 & 550 & $-$ & - & - & - & - & - \\ 
		\hline
		\multicolumn{11}{|c|}{XMM44 [z=1.58]} \\
		\hline
		SN~FM29 & $00:44:02.9$ & $-20:33:50.0$ & 1.6 & 270 & $+$ & $ -1.6 \pm  0.7$ & $-0.06 \pm 0.05$ & $-19.29 \pm 0.06$ & $ 7229 \pm    3$ &   16.93(19) \\ 
		\hline
	\end{tabular}
	\\
	\raggedright{Notes---(1) designation of each SN candidate;(2)(3) J2000 coordinates of the SN, aligned to USNO-B1; (4) angular separation, in arcseconds, between the SN and its host galaxy (if an obvious host exists); (5) projected distance, in kpc, between the SN and approximate center of cluster; (6) decision on whether transient is a SN Ia at the cluster redshift: $+$=yes, $-$=no, ?=possible; (7)-(10) best fitting SALT2 SN~Ia model parameters, and $1\sigma$ uncertainties, for events deemed likely or possible SNe~Ia in their clusters; (11) $\chi^2$ and degrees of freedom of best cluster SN~Ia model fit.}
\end{table*}
Figure~\ref{fig:cluster_superplots} shows the location of each of the transients in their respective cluster fields, with zoom-ins on the transient locations and host galaxies. 
\begin{figure}
	\includegraphics[scale=0.5]{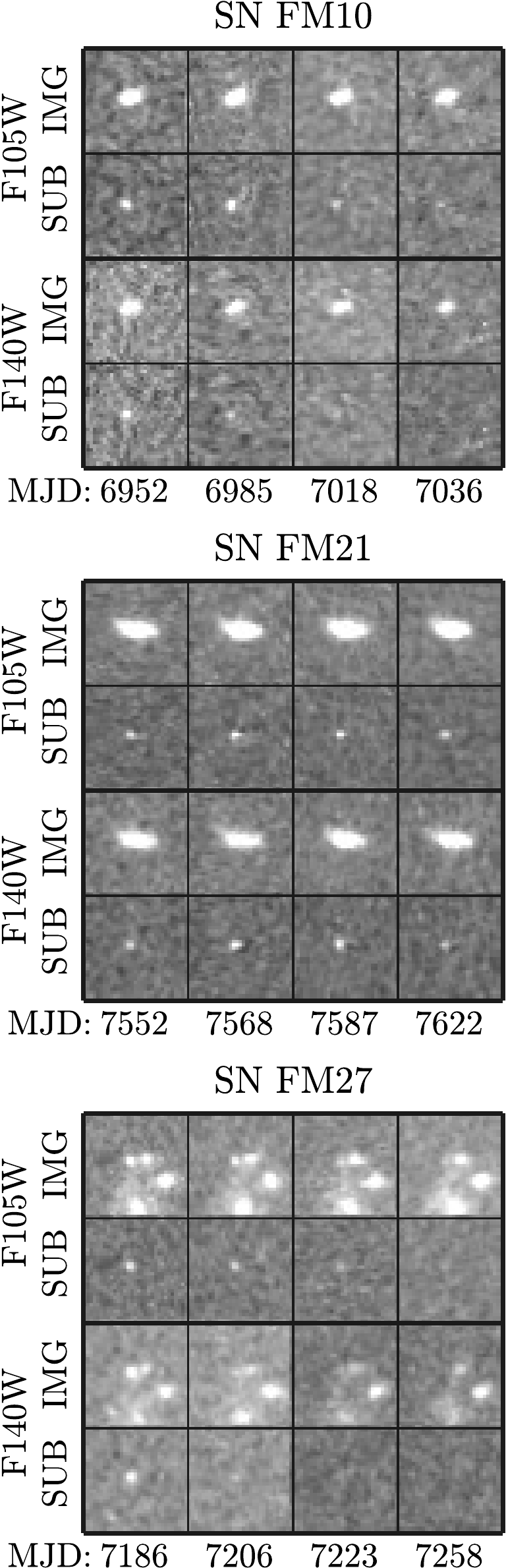}
	\centering
        \caption{Examples of difference images for SN~FM10, SN~FM21 and SN~FM27 in the two main survey bands. The top row for each transient and band is the registered image and the bottom row is the difference image, showing the transient evolving across the observation epochs. Dates for each epoch are in MJD-2450000.}
    \label{fig:sub_images}
\end{figure}
Figure~\ref{fig:sub_images} illustrates, for several of the detected transients,
image sections of the individual epochs and the difference images that reveal the transient.
Figure~\ref{fig:cluster_colormag} is an example of a color-magnitude diagram of the galaxies in a cluster field, with the transient's host galaxy marked, along with any cluster members that have been spectroscopically confirmed as such in the literature. We have produced and examined such a color-magnitude diagram for every cluster field in the sample, and used it as additional input in assessing the likelihood that a transient's host galaxy is a cluster member.  
\begin{figure}
	\includegraphics[width=\columnwidth]{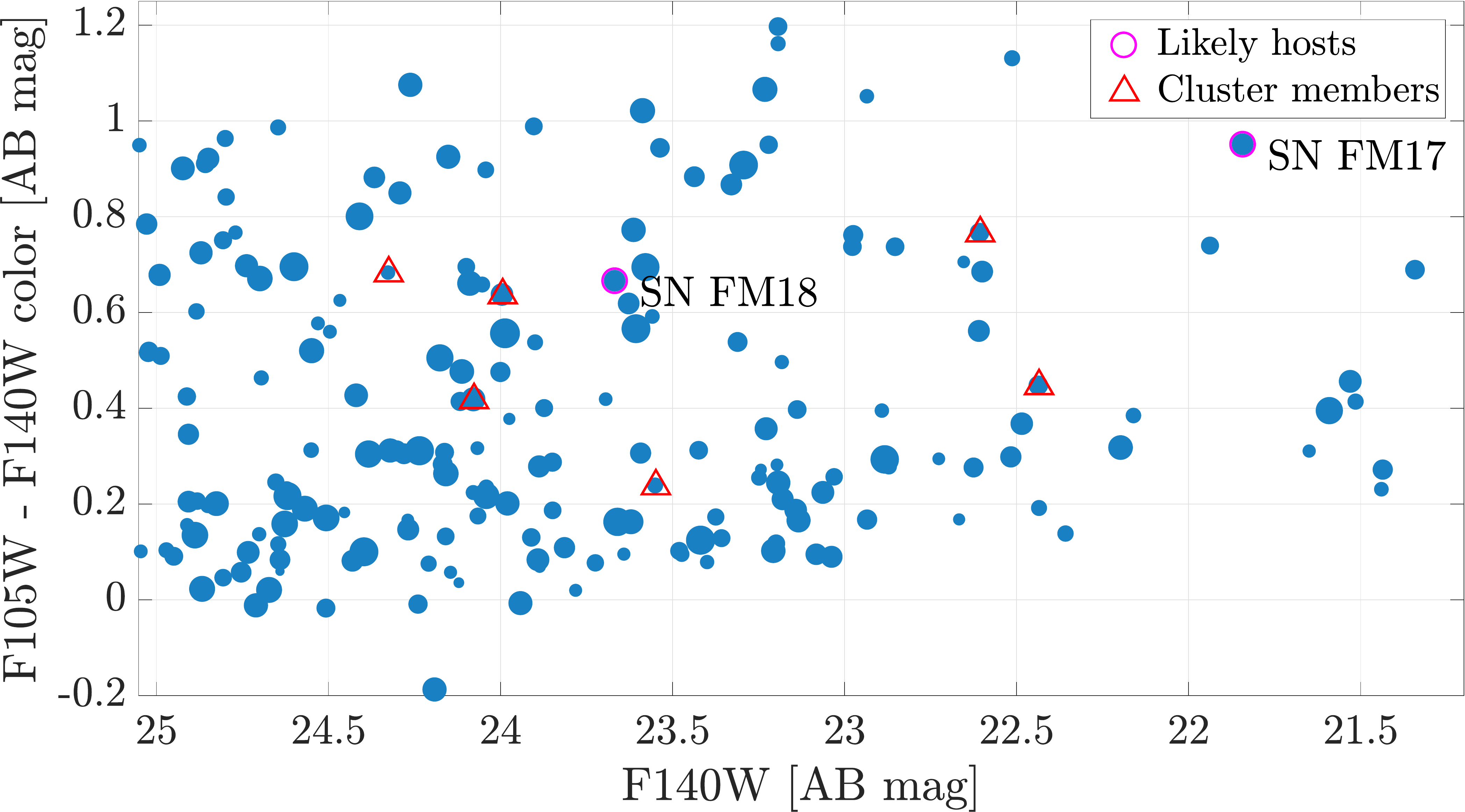}
        \caption{Example of a color-magnitude diagram of all galaxies in a cluster field, for the cluster SPARCS0330. Spectroscopically confirmed cluster members are marked with red triangles. SN host galaxies are marked by magenta circles. The size of the symbols is proportional to a galaxy's projected distance from the center of the cluster.}
    \label{fig:cluster_colormag}
\end{figure}

Light curves for all \TotNumofSNe ~transients in all observed bands are shown in Figure~\ref{fig:sne_superplots}, each light curve
 with its best-fitting (but not necessarily acceptable, see below) model light curve of a SN~Ia at the cluster redshift.
All photometric measurements for our detected transients are given in Table \ref{tab:phot_table}. 

\begin{figure*}
	\includegraphics[width=\textwidth]{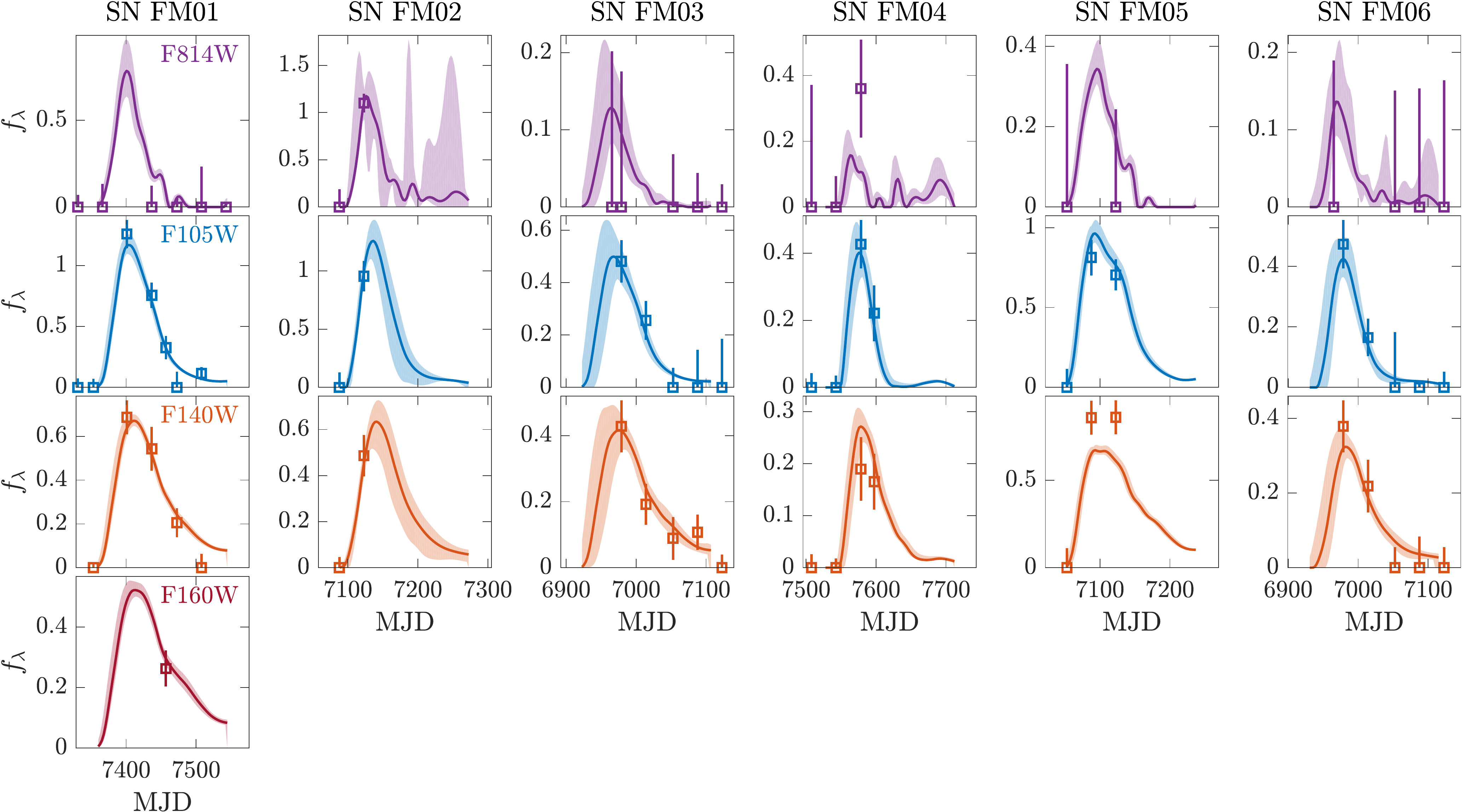}
        \caption{Transient light curves and model fits (not necessarily acceptable, see text) of a SN Ia at the cluster redshift. Shaded regions show the $1\sigma$ ranges around the best-fit model.  $f_{\lambda}$ in units of $10^{-19}~[{\rm erg} \ {\rm cm}^{-2}~ {\rm s}^{-1}{\rm \text{\AA}}^{-1}]$.}
    \label{fig:sne_superplots}
\end{figure*}

\begin{figure*}
	\includegraphics[width=\textwidth]{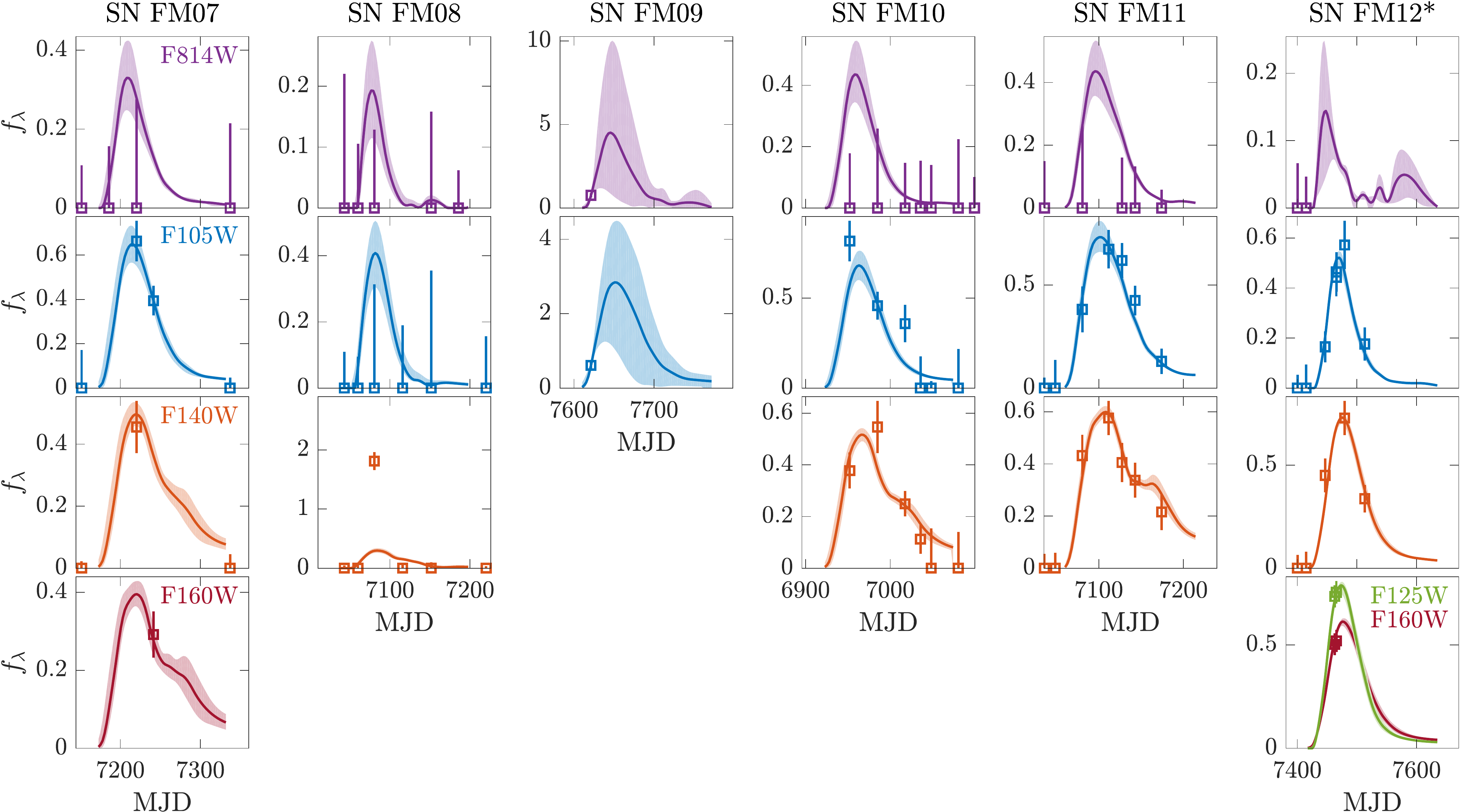}
        \caption{Fig.~\ref{fig:sne_superplots}, continued.}
	\raggedright{* SN~FM12 is fit with a model of a SN~Ia at $z=2.2$, gravitationally magnified by a factor 2.8, as reported in \citealt{Rubin2017} (see text).}
    \label{fig:sne_superplots_2}
\end{figure*}

\begin{figure*}
	\includegraphics[width=\textwidth]{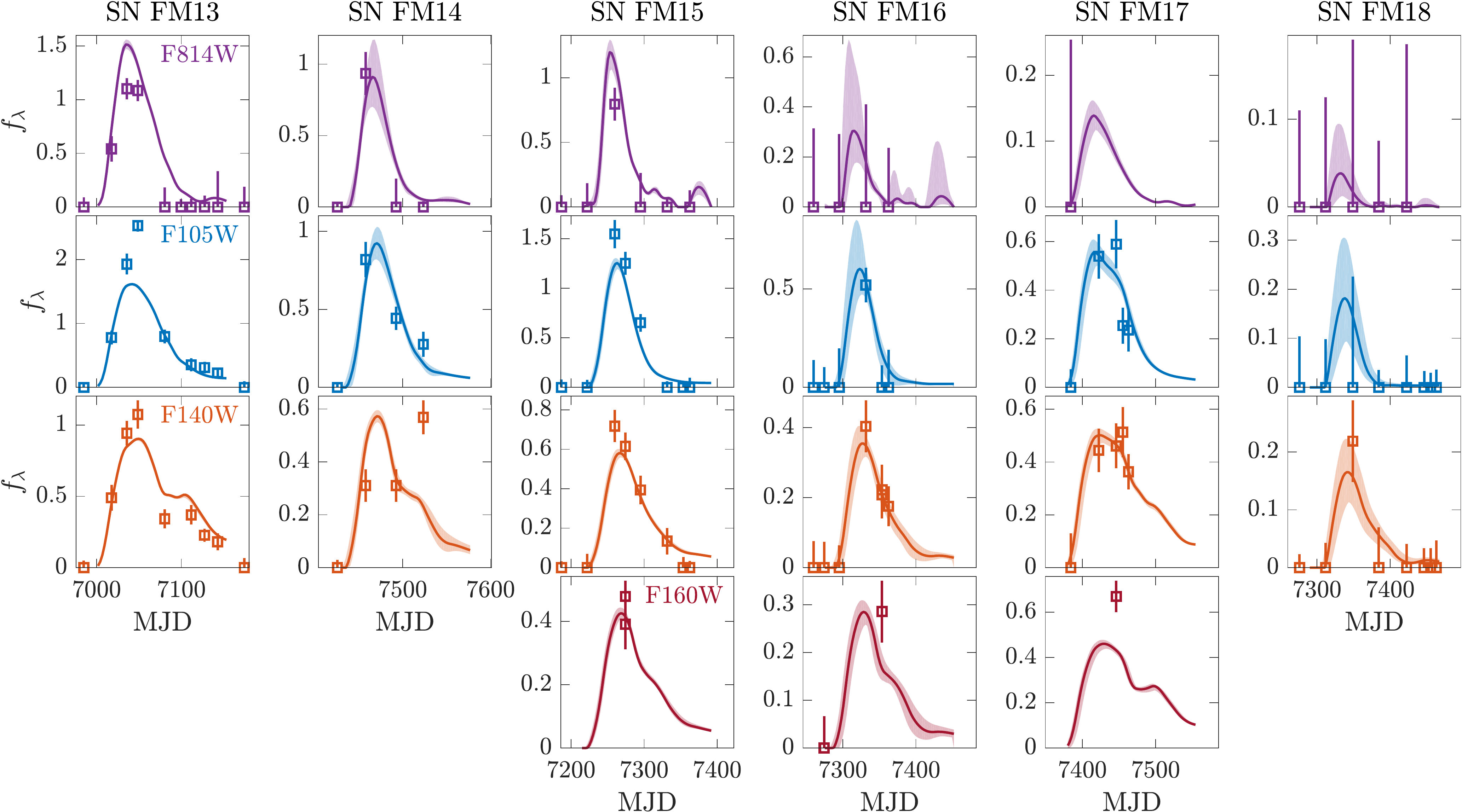}

        \caption{Fig.~\ref{fig:sne_superplots}, continued.}
        
    \label{fig:sne_superplots_3}
\end{figure*}

\begin{figure*}
	\includegraphics[width=\textwidth]{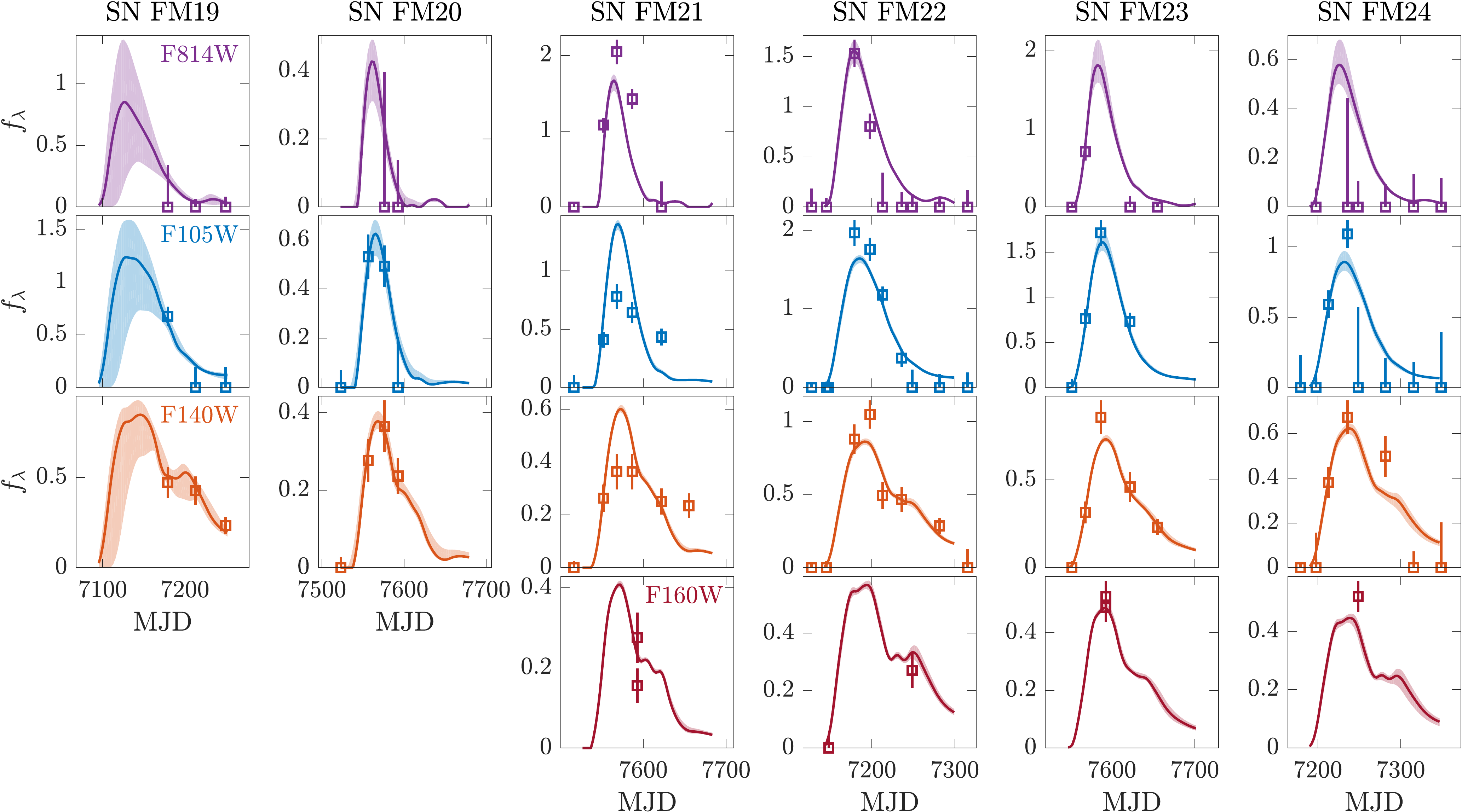}
        
        \caption{Fig.~\ref{fig:sne_superplots}, continued.}

    \label{fig:sne_superplots_4}
\end{figure*}

\begin{figure*}
	\includegraphics[width=\textwidth]{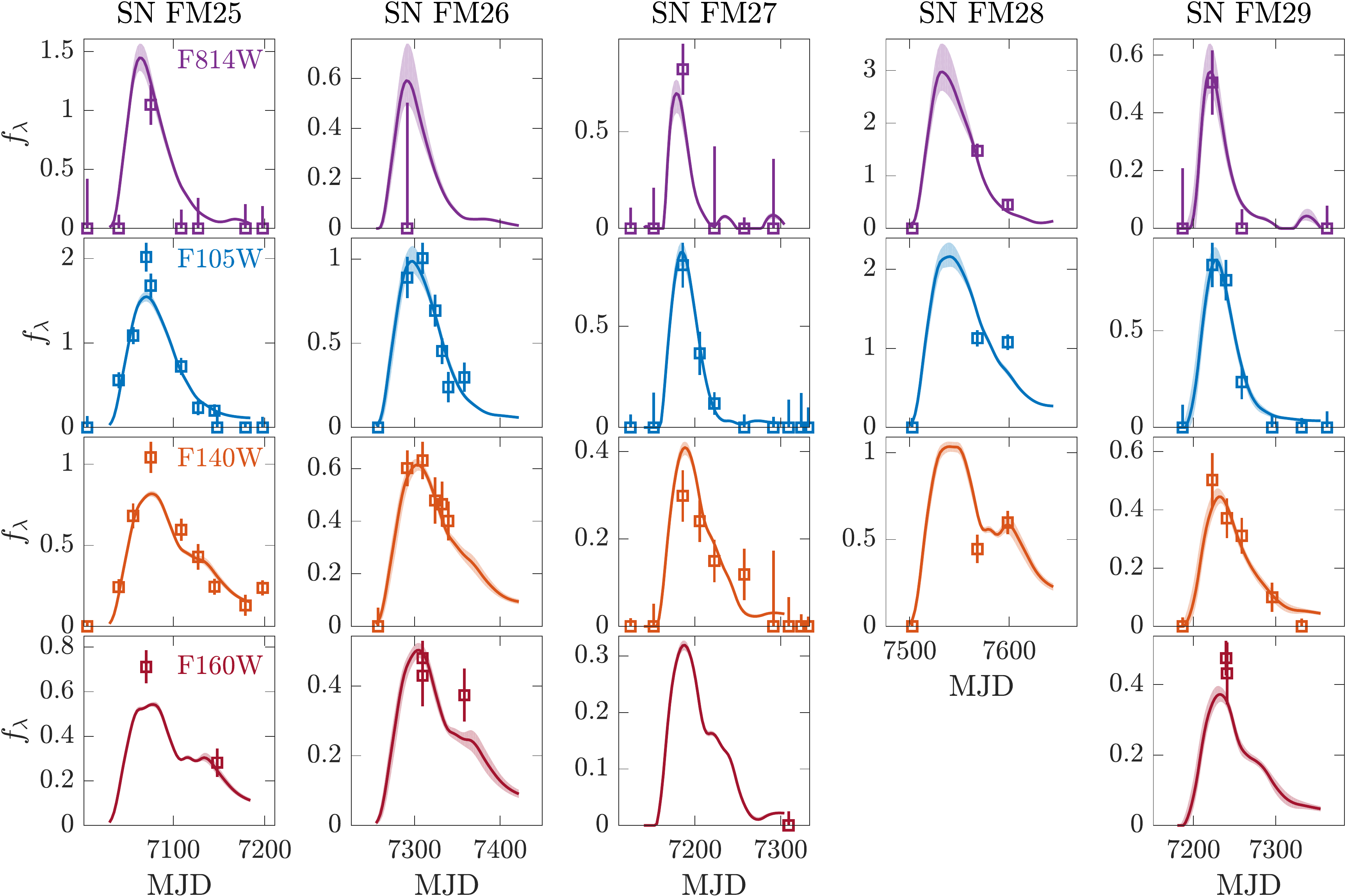}
        
        \caption{Fig.~\ref{fig:sne_superplots}, continued.}

    \label{fig:sne_superplots_5}
\end{figure*}

\begin{table}
	\centering
	\caption{SN candidate photometry}
	\label{tab:phot_table}	\begin{tabular}{cccc}
		\hline
		 MJD & Band & $f_\lambda$ & $\Delta f_\lambda$ \\ 
		\hline
		\hline
		\multicolumn{4}{|l|}{SPARCS3550 [z=1.34]} \\
		\hline
		\hline
		\multicolumn{4}{|l|}{SN~FM19} \\
		\hline
		7179.125 & F105W & 0.674 & 0.093 \\ 
		7179.117 & F140W & 0.472 & 0.087 \\ 
		7212.859 & F140W & 0.427 & 0.079 \\ 
		7249.652 & F140W & 0.234 & 0.046 \\ 
		\hline
		\multicolumn{4}{|l|}{SN~FM20} \\
		\hline
		7556.368 & F105W & 0.532 & 0.090 \\ 
		7576.099 & F105W & 0.493 & 0.084 \\ 
		7556.356 & F140W & 0.277 & 0.055 \\ 
		7576.110 & F140W & 0.365 & 0.067 \\ 
		7592.733 & F140W & 0.237 & 0.047 \\ 
		\hline
		\hline
	\end{tabular}
\\
\raggedright{Notes--MJD values are minus 2450000. $f_\lambda$ values in units of $[10^{-19}~{\rm erg} \ {\rm cm}^{-2} ~{\rm s}^{-1}{\rm \text{\AA}}^{-1}]$. Table \ref{tab:phot_table} is published in its entirety in the electronic edition of Monthly Notices of the Royal Astronomical Society. }
\end{table}

Our objective is to measure the SN~Ia rate in clusters, and hence we need to distinguish SNe~Ia in cluster-member galaxies from all other transients (whether within the clusters, foreground or background to them). We therefore have not attempted to fit the photometry with light curve models of non-Ia SNe.
To fit the observed transient light curves, for every cluster we generate a grid of model SN~Ia light curves at the cluster redshift, in the observed HST bands, using the SNcosmo program\footnote{http://sncosmo.readthedocs.io/} \citep{Barbary2016SNCosmo}. 
The program creates model SN~Ia light curves using the SALT2 \citep{Guy2010} spectral templates. SALT2 models have three interesting parameters: $x_0$, which is an overall normalisation
(related to the peak rest-frame $B$-band apparent magnitude by $m_B = 10.635-2.5~{\rm log_{10}}~x_0$), $x_1$, which quantifies the light-curve width; and the parameter $c$, which expresses a combination of the effects of reddening by dust (quantitatively roughly equivalent to $E(B-V)$) and of the intrinsic color variance of SNe~Ia. 
Adopting the observed distribution of properties of SNe~Ia in massive galaxy clusters, as found by \citet{Xavier2013}, we create a grid of models over the parameter ranges $-4<x_1<4$ and $-0.4<c<0.6$.
  
For every choice of these parameters, the absolute magnitude at maximum light in the rest-frame $B$ band is obtained through a SN~Ia luminosity-width-color relation \citep{Xavier2013},
\begin{equation} \label{mB_eq}
    M_B = M_\star - \alpha x_1 + \beta c, 
\end{equation}
where we assume $M_\star=-19.39$~mag, $\alpha=\XavierAlpha$ and $\beta=\XavierBeta$, again based on the results of \citet{Xavier2013} for SNe~Ia in clusters. 
For every choice of parameters, the time-evolving SN~Ia spectrum is suitably redshifted and folded through the HST system+filter bandpasses. 
The flux density in the bandpass is calculated for a luminosity distance based on a cosmological model (assumed throughout this paper) with $H_0=70 ~{\rm km~s}^{-1}~{\rm Mpc}^{-1}$, $\Omega_{\Lambda}=0.7$, and $\Omega_{\rm m}=0.3$, and including a division or multiplication by a $1+z$ k-correction to obtain $f_\lambda$ or $f_\nu$, respectively. 
The observer-frame time axis is stretched by a $1+z$ cosmological time dilation factor. 
To fit each observed transient light curve, we calculate $\chi^2$ for every model in the two-parameter grid, with each model also running over a range of times of maximum light around an initial guess. 
$1\sigma$ uncertainties around the best-fit $x_1$ and $c$ values are estimated by requiring a change $\Delta \chi^2=1$. 

SALT2 light curves become less reliable at rest-wavelengths below $\sim 3500$~\AA\ \citep{Guy2007,Guy2010} due to poor ultraviolet characterisation of the templates, and therefore in clusters above $z\sim 1.4$, we sometimes examine the effect of excluding F814W data from the fits.
Table~\ref{tab:sne_table} lists for each SN 
our decision on its classification as a SN~Ia in a cluster galaxy, and the best-fit SALT2 model values and uncertainties for the 15 among the \TotNumofSNe~ candidate transients which we deem to be likely or possible SNe~Ia in the surveyed clusters. 
We briefly discuss below each of the \TotNumofSNe ~candidates.

\section{Notes on individual SN candidates}

\subsubsection*{SN~FM01}  
This transient appears far from the cluster center, in a host galaxy with color bluer than the cluster's red sequence. 
A model light curve of a SN~Ia at the cluster redshift fits the data well, although with a rather high luminosity, suggesting possibly a foreground SN Ia. 
In any case, this event is not in this cluster's full-coverage region, and therefore we exclude it from our rate-calculation sample.

\subsubsection*{SN~FM02}
This event appears in three bands but at only one epoch, the last in a season. 
It has no discernible host galaxy. 
Although in principle it could be a SN Ia at the cluster redshift, it is near the field edge, outside the full-coverage area of this cluster, and hence we do not include it in the SN rate sample.

\subsubsection*{SN~FM03}
This declining transient is detected at the beginning of a season, with a clear detection only at two epochs in two bands. 
As such, a model of a SN Ia at the cluster redshift can be fit to the photometry, although if the maximum actually occurred significantly before the first epoch, the
luminosity would be unreasonably high for a SN Ia at the cluster redshift. 
The host galaxy, although at a projected distance not too far from the cluster center ($\sim 260$~kpc) has a spiral morphology and color bluer than the cluster's red sequence. 
We count this event as a possible cluster SN Ia.

\subsubsection*{SN~FM04}
This event is detected in the last two epochs of a season. 
It rises and falls too fast for a SN Ia at the cluster redshift. 
The host is faint and blue. 
We consider this not to be a cluster-member SN~Ia.

\subsubsection*{SN~FM05}
The event is in a host that is projected far from the cluster center and is clearly bluer than the cluster's red sequence. 
The transient is essentially on the galaxy nucleus, and it may even be a subtraction residual artifact on the two epochs in which it is detected, the last of a season. 
The fit to a SN~Ia is poor and requires an extremely slow and luminous, yet extinguished, event, suggesting more likely a foreground Type-IIP SN.
We do not include it in the rate-calculation sample.

\subsubsection*{SN~FM06}
An event in the outskirts of, but clearly associated with, a likely cluster early-type galaxy with colors of the cluster's red sequence. 
The declining event appears in two bands at two epochs and thus peaked before the start of the season. 
The photometry is well fit by a normal SN~Ia at the cluster redshift, and we include it in the sample.

\subsubsection*{SN~FM07}
The transient is in a relatively faint and red galaxy within the cluster's central region, with galaxy color on the cluster red sequence. 
The fit is consistent with a normal, slightly reddened and extinguished SN~Ia at the cluster redshift, and we count it as such. 

\subsubsection*{SN~FM08}
A hostless event near the field edge, outside the full-coverage area and therefore not included in the cluster SN~Ia count. 
It is a peculiar, bright, and fast event that is seen at only one epoch and in only one band, F140W, where it rises and falls within 50 observer-frame days, at most.

\subsubsection{SN~FM09}
A transient projected near the cluster center on a very faint and blue host, and detected on the last two epochs of the season in F105W, on one epoch in F814W and on one in F160W. 
With the few data points, it can be consistent with a SN~Ia at the cluster redshift. 
However, as it was not detected in our “search band” of F140W (no F140W observations were performed when this event was active), we do not count it in our tally of cluster-member SNe. 

\subsubsection*{SN~FM10}
An event in a faint red galaxy in the central region of the cluster, which could be a cluster member. 
The transient likely peaked near the time of the first epoch of the season. 
The fit is consistent with a normal SN Ia at the cluster redshift but with about 1 mag of rest-frame optical extinction.

\subsubsection*{SN~FM11}
This transient is hosted near the center of one of the central cluster early-type galaxies. 
The photometry is fit well with a luminous SN Ia that has undergone about 1.5 mag of extinction. 
We include it in our cluster SN~Ia sample.

\subsubsection*{SN~FM12}
\citet{Rubin2017} have already reported this transient, which our detection pipeline has discovered independently, with our photometry closely matching theirs. 
\citet{Rubin2017} find that this is a normal SN Ia at $z=2.22$ that, together with its host galaxy, is gravitationally lensed by the cluster potential, magnifying the SN flux by a factor 2.8. 
We confirm that such a model fits well the photometry, as can be seen in Fig. \ref{fig:sne_superplots}. 
As this event is not physically in the cluster, we do not include it in our cluster SN Ia sample.

\subsubsection*{SN~FM13}
This is a bright and well-measured event with no detected host galaxy, occurring far ($\sim 420$~kpc in projection) from the cluster center. 
The light curve is poorly fit by a SN~Ia at the cluster redshift, which would in any case by unreasonably luminous at peak. 
We deem this likely to be a foreground non-SN~Ia event.

\subsubsection*{SN~FM14}
This transient occurs far from the cluster center and has no detected host galaxy. 
The light curve is poorly fit by any SN~Ia model, especially as between the last two epochs of observation it shows a doubling in F140W flux simultaneous with a factor-2 decline in F105W, as already noted by \citet{Boone2016}. 
We do not include this event among the cluster SN~Ia events. 

\subsubsection*{SN~FM15}
This transient occurred on the outskirts (2\arcsec ~from the center, corresponding to $\sim 17$~kpc in projection) of a spiral galaxy that is a spectroscopically confirmed cluster member \citep{Nantais2016}.
The photometry is moderately well fit with a SN~Ia at the cluster redshift, although the rise to maximum appears to be too fast, and the color is too blue in F105W$-$F140W (which correspond to 4000~\AA\ and 5300~\AA\, respectively, at the cluster redshift). 
Considering all these facts, we count the event as a possible, but uncertain, cluster SN~Ia.

\subsubsection*{SN~FM16}
This event is hosted by a galaxy near the central cluster region, with colors similar to those of the cluster red sequence, although slightly redder. 
The light curve is well fit by a normal SN~Ia at the cluster redshift, and we include it in the sample.

\subsubsection*{SN~FM17}
This event is hosted close to the center of a galaxy with colors consistent with the cluster's red sequence, and furthermore one whose photometric redshift is consistent with the cluster redshift \citep{Nantais2016}. 
The light curve is not optimally fit with a SN~Ia model, which would suggest a very luminous and slowly evolving event. 
On the other hand, the photometry in this case may be distorted by subtraction residuals from the host galaxy background. 
In view of this, we consider this a possible, but not certain, SN~Ia in a cluster galaxy.

\subsubsection*{SN~FM18} 
This transient appears associated with a disc galaxy with a photometric redshift consistent with the cluster's redshift. The transient is projected $\sim 1\farcs 5$ from the galaxy, corresponding to $\sim 10$~kpc at the cluster redshift. 
However, the transient is detected at only one epoch and only in F140W, and appears to be too fast and too faint for a normal SN~Ia (see also SN~FM08).
We therefore exclude this event from the cluster SN~Ia sample.

\subsubsection*{SN~FM19} 
Although projected not far from the main cluster galaxy concentration, this event has no detected host. 
The declining light curve peaks before the first epoch of the season. 
To be fit as a SN~Ia at the cluster redshift requires an unreasonably luminous and slow event. 
We exclude this transient from the cluster-rate sample.

\subsubsection*{SN~FM20} 
The host galaxy of this transient is projected within the concentration of cluster galaxies, but the host is blue and unlike the cluster's red sequence. 
The light curve evolution is very fast for a SN~Ia, though not impossible. 
We classify this event as uncertain, but possibly a cluster SN~Ia. 

\subsubsection*{SN~FM21} 
An event in a blue disc galaxy near the field edge, far from the main cluster-galaxy concentration. 
The light curve is poorly fit by SN~Ia models, and may be a foreground Type-IIP supernova. 
We exclude it from the cluster sample.

\subsubsection*{SN~FM22} 
One of two separate transients (the other is SN~FM23, see below) that we detect within a single disc galaxy in the course of a year. The galaxy appears to be of similar color as other spectroscopically confirmed cluster members. 
The light curve is reasonably well-fit by a normal SN~Ia at the cluster redshift. A detailed discussion follows in Section \ref{Sec:dual_sn}, below, of the 
possibility that SN~FM22 and SN~FM23 are 
two gravitationally lensed images of a background SN, magnified and multiply imaged by the combination of the galaxy and cluster potentials, and appearing at different epochs because of the lensing time delay. 
Alternatively, we consider the option that they are 
two unrelated SNe~Ia that exploded within 5.5 rest-frame months in the same cluster galaxy.
We conclude that the evidence favours the latter option, and consider both events as likely cluster SNe~Ia.

\subsubsection*{SN~FM23} 
Appearing about 1 observer-frame year after SN~FM22, and projected on a different part of the same disc galaxy, this event is also fairly well fit as a normal SN~Ia at the cluster redshift. 
As already noted, based on the analysis in Section \ref{Sec:dual_sn}, we consider both SN~FM22 and SN~FM23 as likely  SNe~Ia events in the cluster.

\subsubsection*{SN~FM24} 
Hosted by a spectroscopically confirmed cluster-member disc galaxy, the light curve of this transient is fairly well fit with a normal SN~Ia with about 1 magnitude of optical-band extinction. We include it in the cluster SN~Ia tally.

\subsubsection*{SN~FM25} 
Projected near the cluster center, but on a faint and blue galaxy, this transient’s light curve is well fit with a normal SN~Ia at the cluster redshift, especially if the single F814W measurement is excluded from the fit.
We include this event in the cluster-member SN~Ia sample.

\subsubsection*{SN~FM26} 
This event occurs in a faint galaxy near the cluster's central region, with the galaxy color similar to the cluster's red sequence. 
The photometry is fit well with the light curve of a normal SN~Ia at the cluster redshift.

\subsubsection*{SN~FM27} 
This transient appears projected $\sim 200$~kpc from the cluster's central region. The nearest possible host galaxies are several faint blue ones, 4 to 12~kpc from the transient, or a faint reddish galaxy separated by 15~kpc. The light curve evolution is probably too fast to be a SN~Ia at the cluster redshift, and hence we do not count it as such. 

\subsubsection*{SN~FM28} 
This event is near the edge of the field, outside the fully covered area we use for our rate calculation. 
It is associated with a blue compact host. 
At the two end-of-season epochs with detections, the light curve is poorly fit with a SN~Ia. 
We exclude it from the cluster SN~Ia sample.

\subsubsection*{SN~FM29} 
Near the cluster center and projected $\sim 14$~kpc from a red early-type spectroscopically confirmed cluster galaxy, this event is well fit by a normal SN~Ia at the cluster redshift, and we include it in our sample.

\subsection{SN~FM22 and SN~FM23: a physical double or a gravitationally lensed SN?}
\label{Sec:dual_sn}
As already noted above, SN~FM22 and SN~FM23 are two events separated by about 400 observer-frame days that appear on top of the same likely cluster-member galaxy, at positions separated by $0\farcs49$. 
The host galaxy has the morphology of an inclined disc, it has a bright and relatively blue clump off to one side, and shows some red streaks on the disc. 
The two events are adjacent to the bright clump. 
To try to understand better the host galaxy structure, we have subtracted from the image, at the position of the clump, a scaled version of one of the bright central galaxies in the cluster. 
The result is shown in Figure~\ref{fig:dual_sn_fit}.  
Both events are well fit by normal SN~Ia light curves at the cluster redshift, but with different best-fit SN Ia model parameters---the absolute peak magnitudes are consistent, the color parameter $c$ values are somewhat different, but the $x_1$ light curve width is significantly different, with SN~FM22 being wider.  
The question naturally arises whether these were two unrelated SNe~Ia in the same cluster galaxy, or rather multiply lensed and time-delayed images of the same SN~Ia in a galaxy behind the cluster (as seen for a core-collapse SN behind a cluster by \citealt{Kelly2015}). 
We briefly consider here the two options, and conclude that the option of two unrelated cluster SNe~Ia is favoured.

\subsubsection*{Two unrelated SNe}
In the $z=1.32$ rest-frame of the apparent-host cluster galaxy, the time interval between the two SNe is about 0.46 year. 
In the Milky Way, for example, a SN~Ia explodes about once per $\sim 200$ years (see \citealt{Maoz2014}, or consider the number of historical SNe~Ia--at least three--of the previous millenium). 
While this difference in intervals might suggest it unlikely that the two events are unrelated, one should recall that we are dealing with massive galaxies, in a cluster environment (with its high-normalisation DTD, see Section \ref{Sec:Intro}), and at a large look-back time, i.e. at a short delay after star formation. 
From the DTD for SNe~Ia in clusters in \citet{Maoz2017} (see also the DTD derived in Section~\ref{Sec:dtd}, below), the expected SN~Ia rate per unit stellar mass in the host galaxy is about $2\times 10^{-13} {\rm yr}^{-1} {\Msun}^{-1}$. 
By fitting a stellar population synthesis (SPS) model to the integrated F160W light of the host galaxy of SN~FM22 and SN~FM23 (see Section \ref{Sec:Cluster-Mass}, below), we estimate a stellar mass, at formation epoch, of $\sim 2\times 10^{11} \Msun$, and therefore an expected mean interval of 25~yr between SNe~Ia in this galaxy. 
However, one must beware of {\it a posteriori} statistics and recall that the cluster was monitored for about twice as long as the observed interval between SN~FM22 and SN~FM23, i.e. for a rest-frame interval of about 1 year. 
Furthermore, this "double feature" is one that is seen among 12, or so, cluster SNe~Ia that we have discovered. Thus,
even if the probability for seeing two SNe~Ia in a particular host galaxy during the period of the experiment is 4\%, there is a high probability, $1-0.96^{12}=0.39$, for seeing a second SN in the same host after 12 independent "draws". 

\begin{figure}
	\includegraphics[width=\columnwidth]{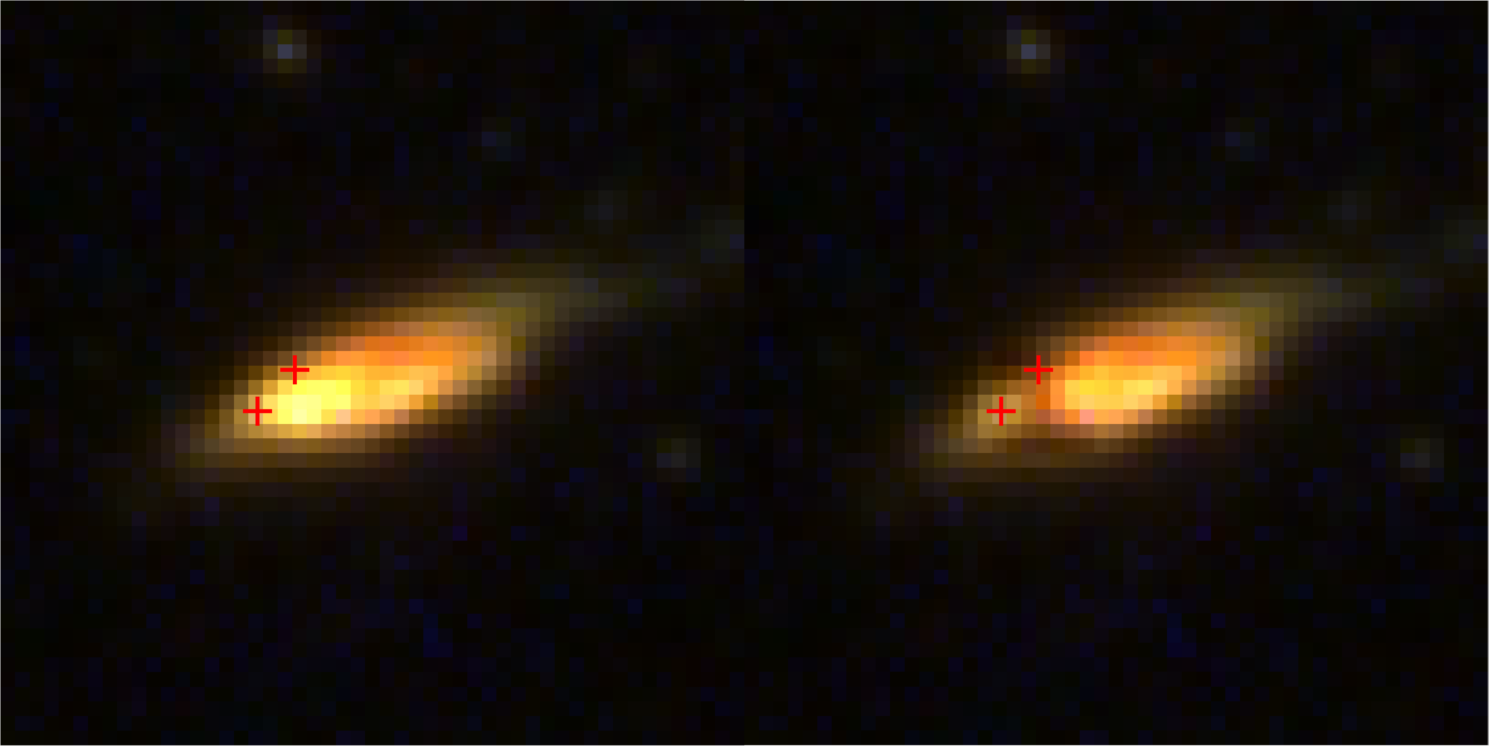}
        \caption{Color rendition of the host galaxy of SN~FM22 and SN~FM23. The left and right panels show the host before and after subtraction of a left-hand-side 'clump', respectively. Red crosses mark the locations of the two transients. }
    \label{fig:dual_sn_host}
\end{figure}

The differences in the model parameters for the two events would be as expected for two unrelated SNe. 
The bright asymmetric clump in the host galaxy could be a recent or on-going star-forming region, and the red streaks could be dust lanes, another possible indicator of star-forming activity. 
The proximity of the SNe to the clump may suggest that they are associated with this young stellar population, which therefore probes a shorter delay of the DTD, resulting in a high SN~Ia rate. Indeed, subtraction of the clump appears to reveal a sub-clump of stars right under the location of SN~FM22 (see Fig.~\ref{fig:dual_sn_host}).

Another possibility under the present option (of two unrelated SNe) is that the host galaxy is a chance superposition of two galaxies---the "clump" and the disc galaxy, which after the subtraction shown in Fig. \ref{fig:dual_sn_host}, appears more symmetric and undisturbed. 
The two galaxies could both be cluster members, or one of them could be background or foreground to the cluster (but not actually strongly lensing each other in this scenario, if the criteria for strong lensing are not satisfied). 
Each of the two SNe could be in a different galaxy among the two galaxies, or both SNe could be in the same galaxy.

\subsubsection*{Two lensed images of the same SN}
\begin{figure}
	\centering
	\includegraphics[width=0.7\columnwidth]{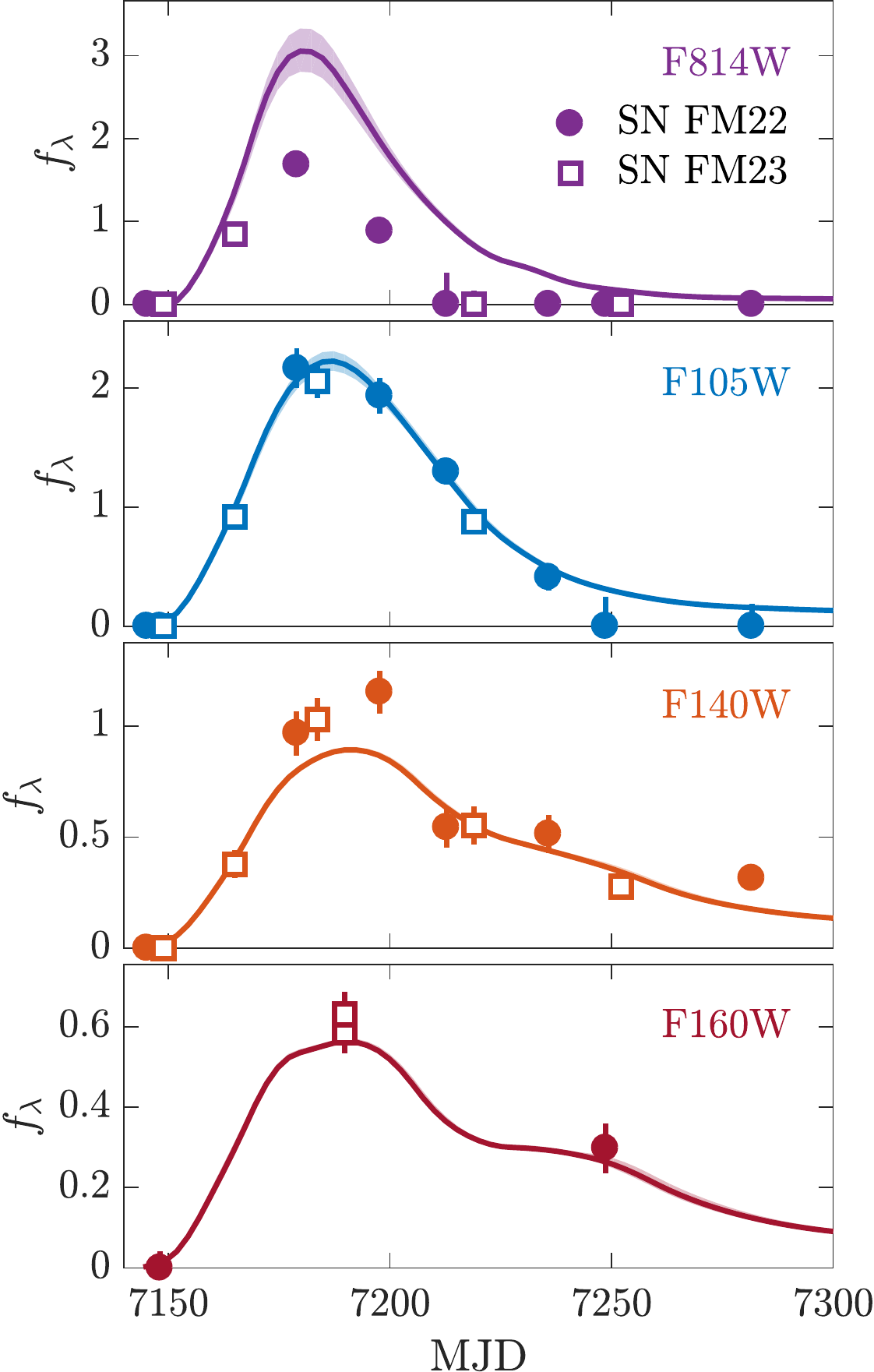}
        \caption{Best-fitting SALT2 model and combined light curves of SN~FM22 and SN~FM23, under the assumption that they are lensed images of a background SN~Ia. The optimal light-curve combination is obtained by varying the time shift between the events, the source redshift, and the possible magnifications of each SN image (see section \ref{Sec:dual_sn}). Flux density in $10^{-19}~ {\rm erg} \ {\rm cm}^{-2}~ {\rm s}^{-1}{\rm \text{\AA}}^{-1}$. Circles and squares represent measurements of SN~FM22 and SN~FM23, respectively.}
    \label{fig:dual_sn_fit}
\end{figure}

To examine the lensing option, we have first attempted to see how well the light curves of the two events can be combined into one, and to what extent this tells us anything about the potentially lensed SN. 
We have fit the SALT2 models, as before, to the two events' light curves in each band, but combining the light curves of each event using four parameters: the time shift between the events, a possible magnification for each event, and the source redshift of the background SN. 
Figure~\ref{fig:dual_sn_fit} shows the best-fit model, which has the source at $z=1.32 \pm 0.08$, an observer-frame delay of $401\pm1$~days between images, and magnifications of 1.1 (for SN~FM22) and 1.2 (for SN~FM23). 
In other words, the best-fit model is for two unlensed and unrelated (though similar) SNe~Ia at the cluster redshift. 
However, there is also a local maximum in the likelihood space for a model with a source at $z \sim 1.7$, but again with magnifications of 1.1 and 1.2 (for SN~FM22 and SN~FM23, respectively). 
Considering the uncertainty of SALT2 models in the rest-frame UV, and to ascertain that more realistic UV models would not allow a source at a higher redshift (which would then be seen even further into the rest-frame UV), we have used the UV-optical SN~Ia spectral models of \citet{Foley2016}, which are based on a sample of observed SNe~Ia with a range of intrinsic properties. 
The SN~Ia colors that the models produce through the HST bands when moving the source to different redshifts, compared to the observed colors of SNe FM22 and FM23, particularly the high observed flux in the bluer among the HST bands, confirm that the SNe can only be at $z\sim 1.32$ (i.e. in the cluster) or at $z\sim 1.7-1.8$. 
We note that the model fit to the combined light curve is poorer than the fits to the individual light curves, with F104W underpredicted and F814W overpredicted, another indication against the lensing option. In the same vein, the significant differences in SALT2 parameter fits to the two individual events is at odds with the achromaticity expected of lensing. 
Although microlensing by stars in the lens galaxy can introduce some chromatic light curve effects starting 3 rest-frame weeks after maximum \citep{GoldsteinNugent2018}, the color differences are probably smaller than seen here. 

We next consider whether there is a plausible lensing configuration that can broadly reproduce the observations under this scenario. The gravitational potential of SPT0205 cannot strongly lens (i.e. lens into multiple images) the two SNe at their observed position. 
The cluster displays a prominent blue lensed arc (at a radius of a few arcseconds from the cluster center, as determined from the center of the innermost SZE contour, see zoom in Figure~\ref{fig:cluster_superplots}). 
The arc roughly marks the location of the lensing critical curve near which multiply lensed images form. 
The two SNe and their underlying galaxy 
are about 50\arcsec~ from the cluster center, i.e. far from the strong-lensing region. 
Nevertheless, the two SNe could be images produced by the lensing effect of the galaxy (or of one or both galaxies if they are a chance alignment of two galaxies, see above), with some additional convergence magnification $\kappa$ from the projected surface mass density of the cluster at this location. 
A $\sim0\farcs5$ image separation is indeed typical for galaxy lenses, but would be too large if the source SN is as close behind the lens (i.e. at relatively low redshift) as the SN colors indicate.
The implied time delay of 400 days is large for a lens system with this image separation, although  the cluster's "mass sheet" convergence could boost the time delay somewhat, by $(1-\kappa)^{-1}$. 

Assuming the main lensing body is the bright clump next to the SNe, the geometry of the images does not correspond to a "double" image configuration, since the two images would straddle the lens, whereas here they are off to one side of the clump. 
If the lens were a four-image "quad" system, the images could form a "diamond" geometry around the lens (e.g. as is \citealt{Kelly2015}) but the time delays between images in such small-separation lenses would be very small, and all four images would have been seen. 
An alternative quad geometry would be with an additional pair of images on the opposite side of the lens, with a larger (smaller) separation from the lens and larger (smaller) separation between the images. 
Presumably those images would have arrived with delays outside the observing window. 

Additional information in lensed-quasar and lensed-SN systems can often be obtained from the lensed image of the background host galaxy. 
The disc galaxy could be the lensed host, magnified by the cluster convergence but with little shear distortion. 
The red streaks could be magnified dust features in the lensed host. 
Alternatively the disc galaxy could be in the cluster after all, and serving as the main lens. 
The red features, which after subtraction of the clump seem to form a nearly closed loop, could be (albeit with some imagination) a distorted Einstein ring formed from the SN host galaxy light. 

From the above discussion, we conclude that the possibility of two unrelated SNe~Ia in the same cluster galaxy is quite likely, while the lensing option is probably too contrived. An optical-through-IR spectrum of the galaxy or galaxies, if it showed combined spectra of galaxies at different lens and source redshifts, could revive the lensing option. 
For the sake of our cluster SN~Ia rate calculation we consider SN~FM22 and SN~FM23 as two unrelated SNe~Ia, both in a cluster galaxy.

\section{Detected SN candidate numbers}
From the discussion above, among the 29 transients we have detected, 11 are consistent with being SNe Ia in the full-coverage areas of their respective clusters, and 4 more are possible cluster SNe Ia. The remaining 14 events are presumably background or foreground or non-Ia SNe. As a check on these numbers, which drive the results of the experiment, we have conducted several tests.

First, we have re-fit SNCosmo SN~Ia light curve models to all 15 likely plus possible cluster SNe~Ia, but leaving the redshift as a free parameter, to see if any of these cases are fit better with a SN~Ia at a redshift other than that of the cluster. In no case was the fit significantly improved by a non-cluster redshift, from which we conclude that it is unlikely that any foreground or background SNe~Ia are contaminating the cluster SN~Ia sample.     

Second, to assess whether the number of non-cluster SNe that we have detected is reasonable, we have made empirical estimates based on two separate field (i.e. non-cluster) SN surveys performed with HST using an observational setup very similar to ours. CANDELS \citep{Rodney2014} used WFC3-IR to search for SNe with 1000~s and 1200~s exposures in the F125W and F160W bands, respectively, similar to the 1000~s F140W exposures in our study. The sum over the product of the number of WFC3-IR fields and the length of time that each field was monitored is 33461 fields$\times$days (where to each continuous monitoring "season" we have added 40 days and 20 days to account for the sensitivity to SNe that peaked before and after the observations, respectively; see Section~\ref{Sec:det-comp}, below). In these data, \citet{Rodney2014} discovered 65 SNe of diverse types and redshifts. In the present work, but considering only the fully covered area of each of our cluster fields, the same type of accounting leads to 2802 fields$\times$days for the cluster SN program.
The expected number of foreground and background SNe in the full-coverage area of the cluster fields is therefore $(65\pm 8)\times 2802/33461=5.4\pm0.7$.

An independent, but less precise, estimate of the expected non-cluster SN count comes from the CLASH survey \citep{Graur2014}, which searched for SNe in the parallel (i.e. non-cluster) fields around rich galaxy clusters. CLASH  again has 1200~s F160W exposures, and 700~s in F125W. Considering the 25 targeted CLASH clusters, each with two parallel WFC3-IR fields, and each field monitored for a median of 54 days, plus the added 40 days and 20 days, as above, leads to  5700 fields$\times$days, in which \citet{Graur2014} discovered $9$ SNe. This then predicts  
$(9\pm 3)\times 2802/5700=4.4\pm1.5$ non-cluster SNe in the present study, nicely consistent with the previous estimate based on CANDELS. 

In our study, among the 14 non-cluster-SN-Ia candidates, 9 are found within the active, full-coverage search area and search band. Among these 9,  SN~FM12 has been gravitationally magnified by the cluster potential by a factor 2.8, so likely would have not been discovered in the absence of the foreground cluster, leaving 8 events. This is an excess of 2-3 events over the CANDELS non-cluster SN prediction, above. At least some of this excess could be from SNe that are in cluster galaxies but are not Type-Ia. For example SN~FM05 and SN~FM21 both look like they could be Type-IIP core-collapse SNe  in star-forming galaxies residing in their clusters. If so, the number of non-cluster-Ia SNe that we have found would be fully consistent with the expectations from CANDELS and CLASH. In any event, the exercise above shows that  we have probably not contaminated our cluster SN~Ia sample with background and foreground events; if anything, we have conservatively undercounted the cluster SNe, and hence we will conservatively underestimate the true cluster SN~Ia rate.

\section{SN detection efficiency and photometric errors}
\label{Sec:det_eff}

To determine the SN detection efficiency of the experiment, we planted simulated point sources with a range of fluxes in the real F140W images, and measured the fraction of simulated sources that we recover in our transient-detection process, as a function of
the source input brightness. 
Thirty such simulated SNe were planted in each epoch of each cluster field. 
In order for the simulated SNe to follow the galaxy light (as do real SNe~Ia, e.g. \citealt{Raskin2009}), the probability for a simulated SN to land on a given pixel was proportional to the pixel's count rate (but excluding bright foreground stars and galaxies, and artifacts such as diffraction spikes around bright stars). 
To assure objectivity and reliability of the simulation, the simulated SNe were planted blindly in the data and searched-for in parallel to the search for real transients in the data. 
The searcher (MF) did not know, until after an entire image was searched, which are the real transients and which are the simulated ones. 

\begin{figure}
	\includegraphics[width=\columnwidth]{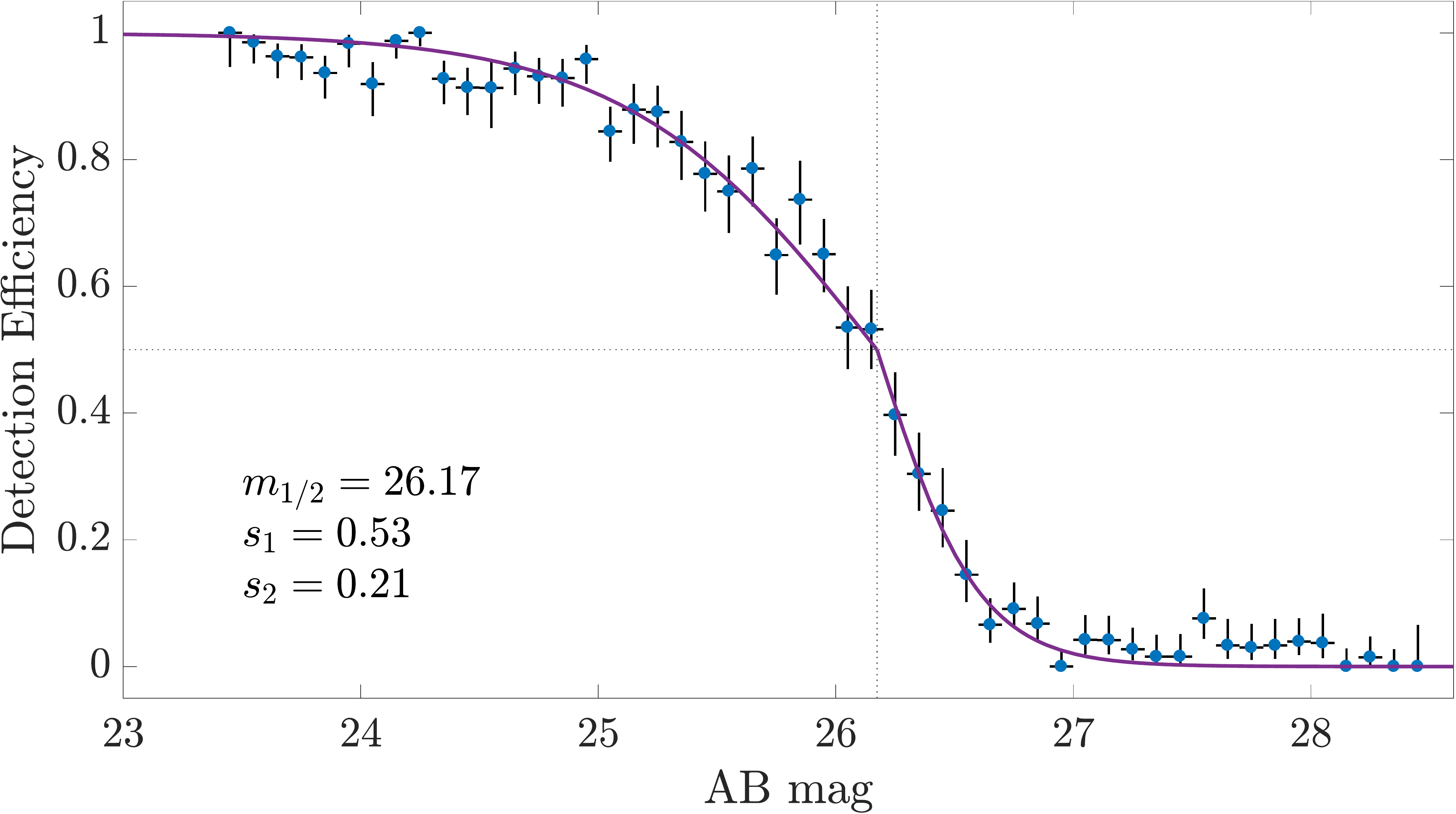}
    \caption{Transient point-source detection efficiency in the F140W band, based on the recovery fraction of simulated sources (see text). The best-fit continuous function of the form of Eq.~\ref{kerenfunction} is also shown.}
    \label{fig:man_det_eff}
\end{figure}

Figure~\ref{fig:man_det_eff} shows the derived transient point-source detection efficiency as a function of simulated source input magnitude. 
The uncertainties on the simulated efficiencies correspond to 68\% binomial-distribution confidence intervals in each magnitude bin. 
Following \citet{Sharon2010} and \citet{Graur2011}, we fit the efficiency measurements with the function
\begin{equation}
\resizebox{0.4 \textwidth}{!} 
{
$
\eta(m;m_{1/2},s_1,s_2)= 
\begin{cases}
    \left( 1+e^{\frac{m-m_{1/2}}{s_1}} \right)^{-1},& m \geq m_{1/2}\\
    \left( 1+e^{\frac{m-m_{1/2}}{s_2}} \right)^{-1},& m<m_{1/2}, 
\end{cases}
$
}
\label{kerenfunction}
\end{equation}
where $m$ is the AB magnitude, $m_{1/2}$ is the magnitude at which the efficiency drops to 50\%, and $s_1$ and $s_2$ determine the range over which the efficiency drops from 100\% to 50\%, and from 50\% to 0, respectively. 
The detection efficiency drops below 50\% at $m_{\rm F140W}\approx 26.2$~mag. 

From a subset of these same simulations, we also determine the photometric precision of our measurements, by performing PSF fitting to the recovered planted transients, as for the real transients, and measuring the variance of the recovered-versus-input fluxes, as a function of input flux. 
Such simulations were performed also on images in the other bands (F814W, F105W, F160W) in order to find the photometric errors in those bands as well. Table~\ref{tab:photometry_stats} summarises the photometric errors, based on these simulations.

\begin{table}
	\centering
	\caption{Photometric uncertainties ($1\sigma$)}
	\label{tab:photometry_stats}
	\begin{tabular}{ccccc}
		\hline
		AB mag & F814W & F105W & F140W & F160W \\ 
		\hline
		23.0-23.5 & 0.05 & 0.03 & 0.07 & 0.02 \\ 
		23.5-24.0 & 0.08 & 0.04 & 0.09 & 0.04 \\ 
		24.0-24.5 & 0.09 & 0.09 & 0.10 & 0.06 \\ 
		24.5-25.0 & 0.08 & 0.10 & 0.12 & 0.11 \\ 
		25.0-25.5 & 0.09 & 0.14 & 0.18 & 0.20 \\ 
		25.5-26.0 & 0.16 & 0.17 & 0.20 & 0.22 \\ 
		26.0-26.5 & 0.22 & 0.28 & 0.30 & 0.26 \\ 
		26.5-27.0 & 0.38 & 0.35 & 0.44 & 0.33 \\ 
		27.0-27.5 & 0.43 & 0.42 & 0.60 & 0.50 \\ 
		27.5-28.0 & 0.62 & 0.57 & 0.90 & 0.76 \\ 
		\hline
	\end{tabular}
\end{table}

\section{SN I\MakeLowercase{a} detection-completeness correction}
\label{Sec:det-comp}

For a proper estimate of a SN rate, one must take into account the incompleteness of the experiment, i.e. the fraction of the cluster SN Ia population that exploded in every cluster field during the survey, but was missed, because of the particular time-sampling pattern of the observations, combined with the detection efficiency as a function of magnitude. 

To obtain the completeness factor of every cluster field, we generated with SNCosmo 10,000 model SN~Ia light curves at the cluster redshift, with times of maximum light  distributed evenly over the observation period. 
The $x_1$ and $c$ parameters of the simulated SNe (with the absolute magnitudes resulting from Eq.~\ref{mB_eq}) were drawn from Gaussian distributions with the mean and root-mean-square (rms) of those distributions for observed SNe~Ia in "passive" cluster galaxies, as measured by \citet{Xavier2013}: $\bar x_1=-0.40$; rms$(x_1)= 1.38$; $\bar c=-0.025$; rms$(c)=0.113$.  
The model light curves were then sampled at the same number and specific times of epochs of the real experiment for a given cluster. 
At each epoch, we noted the detection efficiency for the model SN magnitude, based on the detection-efficiency function (Eq. \ref{kerenfunction} and Fig.~\ref{fig:man_det_eff}). 
The sum of the detection efficiencies at all epochs, divided by the simulated number of SNe, gives the detection completeness factor. 
To avoid multiple counting of the detection probability of the same simulated SN, the detection probability of a single simulated SN was calculated as 1 minus the probability of not detecting it at any epoch: $P_{\rm det}=1-\prod_{i}(1-P_i)$, where $i$ runs over a field's observation epochs.

Following \citet{Graur2014}, we calculated the completeness factor separately for SNe whose maxima occurred {\it before} the first observed epoch of a cluster's observing "season" (which typically lasted about 6-8 months, see Fig.~1), {\it during} the season, and {\it after} the last epoch. As in \citet{Graur2014}, we also use {\it before} and {\it after} periods of 40 and 20 days, respectively, since a SN that reached peak brightness within those time periods could have been bright enough to be detected during the actual observing period. 

Table~\ref{tab:fields_stats} gives the completeness factors for each cluster in the sample. To gauge the systematic uncertainty in the completeness, resulting from the assumptions about the properties of SNe~Ia in clusters, we have re-calculated the completeness, but rather than assuming the distributions of the $x_1$ and $c$ parameters and the values of the $M_B$, $\alpha$, and $\beta$ parameters in passive cluster galaxies, we have replaced them with the parameters for field and "active" galaxies, as measured by \citet{Xavier2013}. This changes the completeness factors in individual clusters, typically by a few percent, but in some cases by up to 25\%. We include this systematic uncertainty in the error budget of our final SN~Ia rate.    

\begin{table*}
	\centering
	\caption{Survey and SN rate input parameters. }
	\label{tab:fields_stats}
	\begin{tabular}{cccccccccccccc}
		\hline
		 Name & $z$ & Delay & N & N & N & $f_{\lambda}$ & $f_{\lambda}$ & $f_{\lambda}$ & $M_{\star}$ & $\tau$ & $\eta_{-40}$ & $\eta_{0}$ & $\eta_{+20}$ \\ 
		  &  & [Gyr] & Ia & Ia? & $\cancel{\rm {Ia}}$ & F105W & F140W & F160W & $[10^{13}~{ \rm M_{\odot}}]$ & [days] &  &  &  \\ 
		 (1) & (2) & (3) & (4) & (5) & (6) & (7) & (8) & (9) & (10) & (11) & (12) & (13) & (14) \\ 
		\hline
		  IDCS1426 & 1.75 & $ 1.5- 2.1$ & 1 & 1 & 4 & $0.27 \pm 0.02$ & $0.27 \pm 0.04$ & $0.33 \pm 0.00$ & $0.52 \pm 0.18$ &   52& 0.71 & 0.86 & 0.75 \\ 
		& & & & & & & & & &   89 & 0.78 & 0.87 & 0.79 \\ 
		  ISCS1432 & 1.40 & $ 2.3- 2.9$ & 1 & 0 & 2 & $0.23 \pm 0.04$ & $0.23 \pm 0.04$ & $0.39 \pm 0.05$ & $0.47 \pm 0.18$ &  104& 0.88 & 0.93 & 0.86 \\ 
		& & & & & & & & & &   87 & 0.88 & 0.96 & 0.86 \\ 
		   MOO1014 & 1.27 & $ 2.7- 3.3$ & 2 & 0 & 2 & $1.68 \pm 0.07$ & $1.47 \pm 0.07$ & $1.47 \pm 0.02$ & $1.45 \pm 0.43$ &   98& 0.93 & 0.99 & 0.89 \\ 
		& & & & & & & & & &   88 & 0.92 & 0.98 & 0.89 \\ 
		   MOO1142 & 1.19 & $ 3.0- 3.6$ & 0 & 0 & 1 & $2.16 \pm 0.06$ & $1.87 \pm 0.16$ & --- & $1.59 \pm 0.55$ &   89& 0.87 & 0.93 & 0.88 \\ 
		SPARCS0224 & 1.63 & $ 1.8- 2.4$ & 1 & 1 & 0 & $0.56 \pm 0.03$ & $0.51 \pm 0.01$ & $0.66 \pm 0.04$ & $0.98 \pm 0.37$ &   67& 0.72 & 0.83 & 0.81 \\ 
		SPARCS0330 & 1.63 & $ 1.8- 2.4$ & 0 & 1 & 1 & $0.26 \pm 0.01$ & $0.22 \pm 0.02$ & $0.34 \pm 0.01$ & $0.48 \pm 0.17$ &  112& 0.77 & 0.87 & 0.85 \\ 
		SPARCS1049 & 1.70 & $ 1.6- 2.2$ & 0 & 0 & 0 & $0.63 \pm 0.00$ & --- & $0.52 \pm 0.08$ & $0.83 \pm 0.38$ &   41& 0.56 & 0.63 & 0.62 \\ 
		SPARCS3550 & 1.34 & $ 2.5- 3.1$ & 0 & 1 & 1 & $0.64 \pm 0.00$ & $0.65 \pm 0.07$ & --- & $0.70 \pm 0.27$ &   75& 0.89 & 0.95 & 0.87 \\ 
		& & & & & & & & & &   45 & 0.88 & 0.97 & 0.88 \\ 
		   SPT0205 & 1.32 & $ 2.6- 3.2$ & 4 & 0 & 1 & $0.50 \pm 0.00$ & $0.46 \pm 0.02$ & $0.58 \pm 0.05$ & $0.63 \pm 0.23$ &  310& 0.93 & 0.99 & 0.88 \\ 
		   SPT2040 & 1.48 & $ 2.1- 2.7$ & 1 & 0 & 1 & $0.95 \pm 0.11$ & $0.79 \pm 0.04$ & $0.96 \pm 0.16$ & $1.27 \pm 0.55$ &  101& 0.85 & 0.96 & 0.91 \\ 
		& & & & & & & & & &   42 & 0.85 & 0.94 & 0.84 \\ 
		   SPT2106 & 1.13 & $ 3.2- 3.7$ & 0 & 0 & 1 & $2.44 \pm 0.06$ & $2.03 \pm 0.12$ & --- & $1.55 \pm 0.49$ &   90& 0.93 & 0.98 & 0.90 \\ 
		& & & & & & & & & &   71 & 0.95 & 0.98 & 0.91 \\ 
		     XMM44 & 1.58 & $ 1.9- 2.5$ & 1 & 0 & 0 & $1.00 \pm 0.11$ & $0.78 \pm 0.02$ & $0.91 \pm 0.05$ & $1.27 \pm 0.46$ &   70& 0.69 & 0.79 & 0.77 \\ 
		\hline
	\end{tabular}
\\ 
\raggedright{
Notes---(1) cluster ID; (2) cluster redshift; (3) delay time range in Gyr, assuming a brief starburst at redshift of $z_f=3-4$; (4)-(6) number of SN candidates in each cluster that are likely cluster SNe~Ia, possible cluster SNe~Ia, and not cluster SNe~Ia, respectively; (7)-(9) total measured galaxy cluster flux within the full-time-coverage area of the field, in the F105W,
 F140W and F160W bands, when available. $f_\lambda$ in $10^{-16} ~{\rm erg} \ {\rm cm}^{-2}~ {\rm s}^{-1}{\rm \text{\AA}}^{-1}$; (10) \textit{formed} cluster stellar mass within the full-time-coverage survey area (see text); (11) cluster rest-frame monitoring time for each season, in days; (12)-(14) detection-completeness factors for the periods before, during and after each observing season, respectively. When more than one season to a cluster, the second season's completeness factors are listed in a separate row.}
\end{table*}

\section{Cluster stellar light and mass estimate}
\label{Sec:Cluster-Mass}

Modern cluster SN rates are normalised relative to the stellar mass of the host galaxy population. 
We estimate this mass by measuring the light in cluster-member galaxies and then translating the light into a stellar mass at the time of cluster formation, using SPS models. 

Measuring the light from cluster-member galaxies is challenging because, in each cluster field, the majority of galaxies are foreground or background to the cluster, and only a few if any of the galaxies (and then only the brightest ones) have been spectroscopically confirmed as cluster members. 
We therefore adopt the approach of \citet{Sharon2010}---statistical background-and-foreground subtraction of non-cluster galaxy light from the cluster images. 
In one implementation of this approach, based on individual galaxy identification, we identify all the galaxies in an image using \textsc{SExtractor}\footnote{https://www.astromatic.net/software/sextractor} \citep{SextractorRef} for source identification, followed by aperture photometry of the identified sources above some local "sky" level. 
The local sky level might include residual instrumental counts (dark current, residual cosmic-ray events), airglow and zodiacal light, unresolved galaxies, intracluster stellar light, and integrated  cosmic background light. 
Galaxies are identified and measured down to the faintest sources that are reliably detected and photometered in a given cluster-field image, and excluding obvious bright foreground stars and galaxies (see cyan circles in Fig.~\ref{fig:cluster_superplots}). 
This is done both for each cluster-field image, and for a non-cluster reference field, observed with the same HST camera setup and filter. 
As a reference field, we have used images from the Hubble Ultra-Deep Field \citep{Beckwith2006} for bands F140W and F160W. 
In the reference field, even though it is a deeper exposure than those of the cluster fields, galaxies are selected only down to the same limiting flux as in the cluster field.  
Assuming the reference field is representative of the field galaxy density and flux in non-cluster directions on the sky, the difference between the total light in galaxies in the two images gives a statistical estimate of the net cluster-galaxy light in the cluster image, above the individual galaxy-identification limit. 

To correct this estimate for the missing light from faint cluster galaxies below the low-luminosity limit, $L_{\rm lim}$, we calculate the fraction of the total luminosity that is missed, assuming the cluster galaxies follow a \citet{Schechter1976} luminosity function, $\Phi (L) dL \propto (L/L^*)^\alpha~ {\rm exp}(-L/L^*)~ d(L/L^*)$, with $\alpha=-1$ \citep{Goto2002}. The correction factor is then 
 \begin{equation}
	C=\frac{\int_{0}^{\infty} L \Phi (L) dL}{\int_{L_{\rm lim}}^{\infty} L \Phi (L) dL}={\rm exp}({L_{\rm lim}/L^*})\approx 1+L_{\rm lim}/L^* .
\label{lumcorr}
\end{equation}
To estimate $L_{\rm lim}/L^*$ in each cluster field, we take the ratio between the limiting galaxy flux and the median flux of the five brightest cluster galaxies, after excluding the brightest two.  
The typical correction factors are small, $C\approx 1.03$.
 
A possible shortcoming of the approach described above, for statistical cluster-light integration and non-cluster-light subtraction, is that it excludes the potential contribution of intracluster starlight to the cluster's stellar mass inventory. 
In low-redshift clusters, intracluster stars may constitute tens of percents of the total stellar mass (e.g. \citealt{Gal-Yam2003, JimenezTeja2018}), although at the high redshifts of the clusters in the current sample, tidal dispersion of the stars from the galaxies into the intracluster medium may have not occurred yet. 
As an alternative, we therefore apply a second implementation of statistical total-cluster-light photometry: we calculate the difference between the total pixel count rates, summed over all pixels, between the cluster-field image and the reference image, regardless of whether the pixel is in a galaxy or not. 
As before, we exclude from both images regions with artifacts, or bright foreground stars or galaxies, properly accounting in the arithmetic for the excluded areas. 
In this method, intracluster light is naturally included, and there is no need for a correction (Eq.~\ref{lumcorr}) to account for faint, uncounted, cluster galaxies. 
However, this method is susceptible to a different systematic error, due to the possibility of a difference in instrumental and near-sky (airglow, zodiacal light) foregrounds between the cluster-field and the reference-field exposures. 

To account for these possible systematics, we have performed the statistical cluster light integration in both of the above ways, with the all-pixel summation typically giving results larger by $\sim 10\%$ compared to the approach of using galaxy identification and aperture photometry. 
We use the range of results we find with the two methods, for every cluster field, as the systematic uncertainty range of this measurement. 
Table~\ref{tab:fields_stats} lists for every cluster field the total stellar flux we thus measure within the monitored cluster area (i.e. the region of each field with full temporal coverage), in each of the three main infrared bands.

To translate the stellar light measurement of each cluster field into an estimate of the formed stellar mass in the monitored field (which is the quantity needed for calculating the DTD), we have used the publicly available Starburst99 SPS program\footnote{http://www.stsci.edu/science/starburst99/} \citep{Leitherer2014} to calculate the spectral evolution, over a Hubble time, of a unit stellar mass of a simple stellar population (SSP, i.e. a single brief burst of star formation at time $t=0$). 
We find the stellar mass by comparing a cluster's total stellar luminosity in some observed band to the luminosity of the SSP in that observed band at an age corresponding to the time interval between cluster star formation and the cosmic epoch at the cluster redshift. 
The redder the observed band, the milder the luminosity evolution of a stellar population and the less sensitivity to extinction by dust, and so we perform our mass-to-light translation using the reddest band generally available, F160W or F140W, which corresponds to rest-frame $V$ to $R$  at the various cluster redshifts. 
We thus minimize the systematic errors due to the assumed redshift of star formation and the detailed star-formation history (which is likely more complex than a SSP; indeed, several of the clusters include galaxies with traces of ongoing star formation, e.g. \citealt{Webb2015}).
For the Starburst99 model calculation we choose a 
\cite{Kroupa2001} IMF (to obtain DTD values compatible with previous DTD estimates that assume this IMF, see Section~1), and Padova+AGB stellar evolution tracks \citep{Girardi2000}. We explore the effect on our results of choosing a range of SPS model metallicities, ${\rm Z}=0.008,0.020$, and $0.050$ (i.e. sub-solar, solar, and super-solar) which systematically affects the derived stellar mass at typical levels of $\pm (30-45)\%$. 

Recent analyses of cluster galaxies, including high-$z$ clusters such as those studied here, confirm that their stars formed in short bursts at redshifts $z_f=3-4$ \citep{Stalder2013,Andreon2016}. 
This range in the possible redshift of star formation, and hence in the age of a cluster, translates into another systematic uncertainty in the stellar-mass-to-light ratio, and therefore in the stellar mass, of $\pm \sim 25\%$. We combine the above systematic uncertainties: in total stellar light (based on the two different statistical cluster-starlight photometry approaches); in mass-to-light ratio, due to assumed model metallicity; and in mass-to-light ratio, due to the assumed star-formation redshift, and hence the stellar age. We thus obtain the total systematic uncertainty 
on our formed stellar mass estimate for every cluster, as listed in Table~\ref{tab:fields_stats}.

\section{SN~I\MakeLowercase{a} rate and delay-time distribution}
\label{Sec:dtd}

We now combine the various elements that we have measured above into an estimate of the SN~Ia rate per formed stellar mass in the HST high-$z$ galaxy-cluster sample. 
The SN~Ia rate is calculated as 
\begin{equation}
 R_{\rm Ia, m*}=\frac{ N_{\rm Ia}}{\sum_i M_i \eta_i \tau_i} ,
\end{equation}
where $N_{\rm Ia}$ is the number of transients that we have classified as SNe~Ia that are within their respective galaxy clusters. Since we have found \NumofSurveyIas ~likely cluster SNe~Ia and \NumofSurveyPossibleIas ~additional possible SNe~Ia, we adopt $N_{\rm Ia}= \TotSNeRateCount \pm \TotSNeRateCountSysErr $~(systematic~error)$\pm \TotSNeRateCountStatErr $~(Poisson error). $M_i$ is the formed stellar mass within the full-coverage monitored area of the $i$th cluster, $\eta_i$ is the SN~Ia detection completeness of a cluster, $\tau_i$ is the "season" monitoring duration of each cluster in the cluster's rest-frame (i.e. the observer-frame duration divided by $1+z$), and the summation is over all of the individual cluster seasons. 
Furthermore, for each cluster season the product $\eta_i \tau_i$ is formed from the sum of the products of completeness and monitoring time for SNe with peaks before, during, and after the actual observing seasons (see Section~\ref{Sec:det-comp}, above). 
To estimate the uncertainty in the SN rate, we combine the Poisson and systematic uncertainties in $N_{\rm Ia}$ and the systematic uncertainties in all of the other parameters, conservatively adopting the extremes of the systematic variations in the rate that can arise by varying all of the parameters within their systematic ranges. 

We obtain a cluster SN~Ia rate of $R_{\rm Ia, m*}=\SurveyIaRate ^{+ \SurveyIaRateUpErr} _{- \SurveyIaRateDnErr} \times 10^{-13} {\rm ~yr}^{-1} {\rm M}_\odot^{-1}$. 
The mean redshift of the cluster sample, weighted by the effectively monitored stellar mass of each cluster, $ M_i \eta_i \tau_i$, is $\langle z \rangle=\SurveyRedshift$. 
Considering the uncertainty in cluster formation redshift, $3<z_f<4$, and the range of cluster redshifts in the sample, the SN Ia rate we measure probes the DTD in the range of delays from \SurveyDelayMin ~to \SurveyDelayMax ~Gyr.

Alternatively, we can split the sample into two sub-samples, one composed of six low-$z$ clusters (at $z=1.13-1.40$), and one with six high-$z$ clusters (at $z=1.48-1.75$), hosting $ \TotSNeRateCountLowz \pm \TotSNeRateCountLowzSysErr $ and $ \TotSNeRateCountHighz \pm \TotSNeRateCountHighzSysErr $ SNe~Ia, respectively.
The resulting rates are then $\SurveyIaRateLowz ^{+ \SurveyIaRateLowzUpErr} _{- \SurveyIaRateLowzDnErr} \times 10^{-13} {\rm ~yr}^{-1} {\rm M}_\odot^{-1} $ at $\langle z \rangle=\SurveyRedshiftLowz$, probing delays of \SurveyDelayLowzMin$-$\SurveyDelayLowzMax~Gyr, and $\SurveyIaRateHighz ^{+ \SurveyIaRateHighzUpErr} _{- \SurveyIaRateHighzDnErr} \times 10^{-13} {\rm ~yr}^{-1} {\rm M}_\odot^{-1}$ at $\langle z \rangle=\SurveyRedshiftHighz$, for delays of \SurveyDelayHighzMin$-$\SurveyDelayHighzMax~Gyr.

\begin{table}
	\centering
	\caption{SN Ia DTD measurements}
	\label{tab:dtd_mes}
	\begin{tabular}{ccccc}
		\hline
		 Redshift & Delay & $R_{\rm Ia, m*}$ & $R_{\rm Ia, L_B}$ & Source \\ 
		  & ${\rm [Gyr]}$ & (DTD) &  & \\ 
		 (1) & (2) & (3) & (4) & (5) \\ 
		\hline
		\multicolumn{5}{|c|}{This work, single redshift bin} \\
		\hline
		$1.35 ^{+0.40} _{-0.22}$ & $ 2.5 ^{+1.2} _{-1.0}$ & $26 ^{+32} _{-15}$ & $0.54 ^{+0.28} _{-0.26}$ & FM18 \\ 
		\hline
		\multicolumn{5}{|c|}{This work, two redshift bins} \\
		\hline
		$1.58 ^{+0.17} _{-0.11}$ & $ 1.9 ^{+0.8} _{-0.4}$ & $35 ^{+66} _{-28}$ & $0.65 ^{+0.61} _{-0.49}$ & FM18 \\ 
		$1.25 ^{+0.15} _{-0.12}$ & $ 2.8 ^{+1.0} _{-0.4}$ & $22 ^{+26} _{-13}$ & $0.48 ^{+0.22} _{-0.21}$ & FM18 \\ 
		\hline
		\multicolumn{5}{|c|}{Previous measurements} \\
		\hline
		$1.12 ^{+0.33} _{-0.22}$ & $ 3.2 ^{+0.9} _{-1.0}$ & $20 ^{+18} _{-15}$ & $0.50 ^{+0.33} _{-0.28}$ & B12 \\ 
		$0.90 ^{+0.37} _{-0.07}$ & $ 4.1 ^{+0.3} _{-1.4}$ & $12 ^{+13} _{-6}$ & $0.80 ^{+1.06} _{-0.52}$ & GY02 \\ 
		$0.60 ^{+0.29} _{-0.10}$ & $ 5.7 ^{+0.7} _{-1.5}$ & $ 8.0 ^{+ 7.2} _{- 6.1}$ & $0.35 ^{+0.31} _{-0.26}$ & S10 \\ 
		$0.46 ^{+0.14} _{-0.26}$ & $ 6.6 ^{+2.3} _{-1.0}$ & $ 9.3 ^{+ 7.2} _{- 6.1}$ & $0.31 ^{+0.51} _{-0.16}$ & G08 \\ 
		$0.25 ^{+0.12} _{-0.07}$ & $ 8.4 ^{+0.7} _{-1.1}$ & $ 5.7 ^{+ 8.3} _{- 3.7}$ & $0.39 ^{+0.90} _{-0.32}$ & GY02 \\ 
		$0.23 ^{+0.75} _{-0.12}$ & $ 8.7 ^{+1.3} _{-0.7}$ & $ 4.5 ^{+ 1.3} _{- 1.0}$ & $0.33 ^{+0.09} _{-0.08}$ & D10 \\ 
		$0.15 ^{+0.04} _{-0.09}$ & $ 9.5 ^{+1.0} _{-0.4}$ & $ 5.0 ^{+ 3.4} _{- 2.5}$ & $0.36 ^{+0.24} _{-0.16}$ & S07 \\ 
		$0.08 ^{+0.08} _{-0.05}$ & $10.3 ^{+0.6} _{-1.0}$ & $ 3.0 ^{+ 1.5} _{- 1.1}$ & $0.23 ^{+0.11} _{-0.08}$ & D10 \\ 
		$0.02 ^{+0.02} _{-0.01}$ & $11.1 ^{+0.2} _{-0.3}$ & $ 3.3 ^{+ 1.4} _{- 1.0}$ & $0.28 ^{+0.11} _{-0.08}$ & M08 \\ 
		\hline
	\end{tabular}
\\
\raggedright{
Notes--- (1) sample redshift; (2) delay times in Gyr; (3) SN~Ia rate per unit formed stellar mass in $10^{-14}~{\rm yr^{-1} M_{\odot}^{-1}}$; (4) SN~Ia rate per unit $B$-band stellar luminosity in $10^{-12}~{{\rm yr}^{-1} L_{B,\odot}^{-1}}$; (5) references for the different measurements: B12--\citet{Barbary2012}; GY02--\citet{Gal-Yam2002}; S10--\citet{Sharon2010}; G08--\citet{Graham2008}; D10--\citet{Dilday2010}; S07--\citet{Sharon2007,Gal-Yam2008}; M08--\citet{Mannucci2008}; }
\end{table}

\begin{figure*}
	\includegraphics[width=0.7\textwidth]{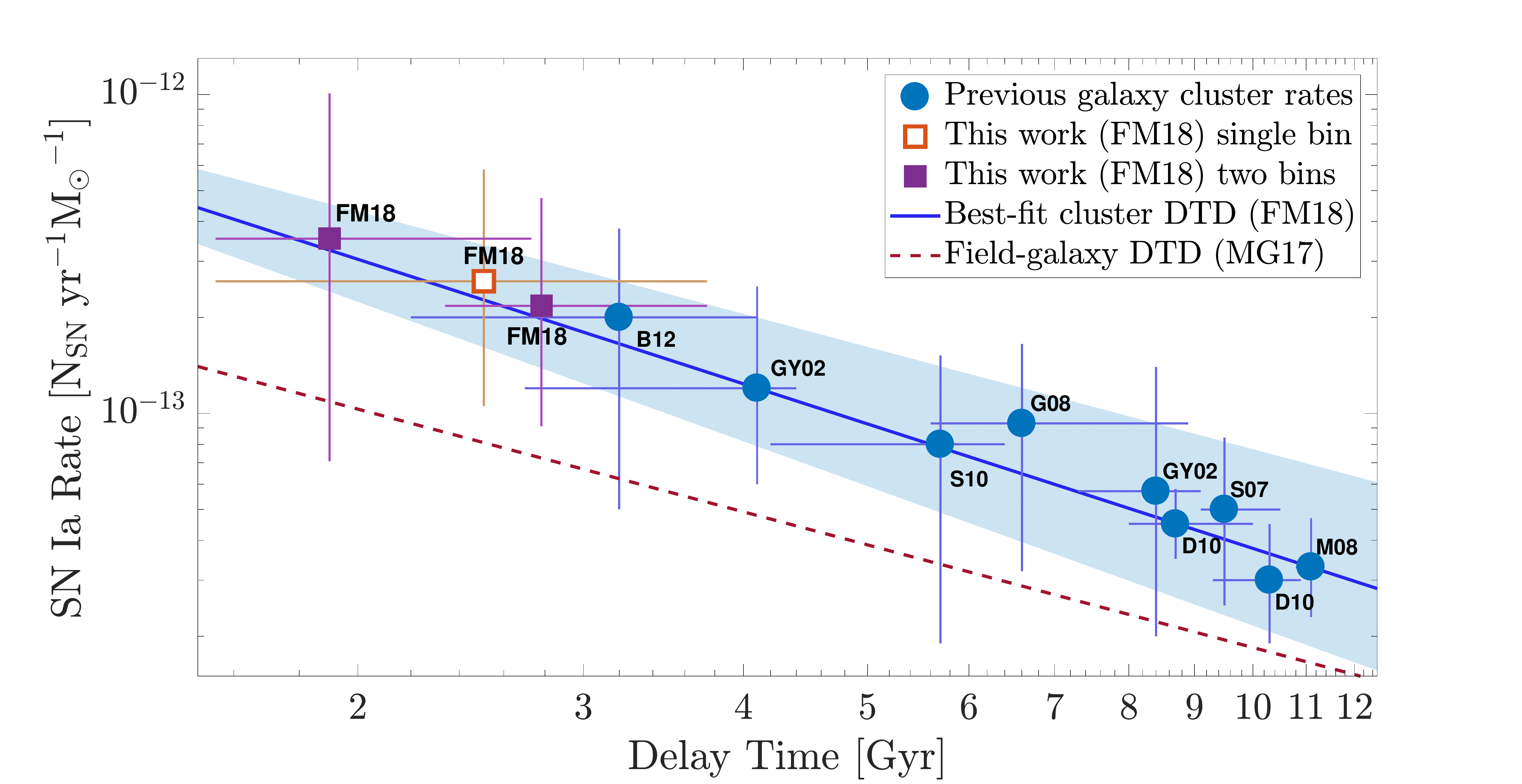}
    \caption{SN~Ia rate measurements, plotted along with previous cluster SN~Ia rates. The empty and filled squares show the single-redshift-bin and two-redshift-bin measurements, respectively. The power-law DTD shown (solid line) is a fit to the two-bin rates, constrained to the $N/M_\star$ normalisation of \citet{Maoz2017}. Dashed line shows, for reference, the field-galaxy SN~Ia DTD from \citet{Maoz2017}. See Table~\ref{tab:dtd_mes} for labels of previous measurements.}
    \label{fig:dtd_final}
\end{figure*}

Table~\ref{tab:dtd_mes} lists,
and Figure~\ref{fig:dtd_final} shows, our new cluster SN~Ia rate measurements alongside previous measurements of the SN~Ia DTD in cluster environments, as recently compiled in \citet{Maoz2017}. 
A two-parameter $t^\alpha$ power-law fit (parameters are power-law index and normalisation) to the DTD measurements with a single cluster-redshift bin gives a best-fit index $\alpha= \AlphaTwoParamOneBin ^{+ \AlphaTwoParamOneBinUErr } _{- \AlphaTwoParamOneBinLErr }$, with a time-integrated (between 40~Myr and 13.7~Gyr) SN~Ia yield of $N_{\rm Ia}/M_\star= \NMsOneBin ^{+ \NMsOneBinUErr } _{- \NMsOneBinLErr } {\rm yr}^{-1} {\rm M_\odot}^{-1}$. If fitting just the power-law index while constraining the integrated number of SNe~Ia to $5.4 \pm 2.3$, based on the total observed iron-to-stellar mass ratio in clusters (see \citealt{Maoz2010, Maoz2017}), then the best fit index is $\alpha = \AlphaOneParamOneBin ^{+ \AlphaOneParamOneBinUErr } _{- \AlphaOneParamOneBinLErr }$. 
Repeating this with the rate measurements in two cluster redshift bins gives $\alpha = \AlphaTwoParamTwoBin ^{+ \AlphaTwoParamTwoBinUErr } _{- \AlphaTwoParamTwoBinLErr }$ and $N_{\rm Ia}/M_\star= \NMsTwoBin ^{+ \NMsTwoBinUErr } _{- \NMsTwoBinLErr } {\rm yr}^{-1} {\rm M_\odot}^{-1}$ (two-parameter fit) and $\alpha= \AlphaOneParamTwoBin ^{+ \AlphaOneParamTwoBinUErr } _{- \AlphaOneParamTwoBinLErr }$ (one-parameter fit). 
The best-fit parameters to the cluster-environment DTD are summarised in Table~\ref{tab:dtd_fit}, along with the best-fit parameters for the field-galaxy SN~Ia DTD from \citet{Maoz2017}.
For convenience, equivalently to $N/M_*$ (the time-integrated number of SNe~Ia per formed stellar mass exploding between 40~Myr and 13.7~Gyr), we list also the DTD normalisation in terms of $\Psi_{\rm 1}$, the DTD amplitude at a delay of 1~Gyr.

\begin{table}
	\centering
	\caption{SN Ia DTD fit parameters. }
	\label{tab:dtd_fit}
	\begin{tabular}{ccccc}
		\hline
		 ${\rm N_{bins}}$ & ${\rm N_{params}}$ & $t^{\alpha}$ & $N/M_{\star}$ & $\Psi_{\rm 1}$ \\ 
		 (1) & (2) & (3) & (4) & (5) \\ 
		\hline
		1 & 2 & $-1.45 ^{+0.51} _{-0.38}$ & $9.0 ^{+29.2} _{-7.1}$ & $1.0 ^{+1.2} _{-0.7}$ \\ 
		2 & 2 & $-1.42 ^{+0.46} _{-0.34}$ & $8.1 ^{+21.0} _{-6.0}$ & $1.0 ^{+1.0} _{-0.6}$ \\ 
		1 & 1 & $-1.30 ^{+0.23} _{-0.16}$ & $5.4 ^{+2.3} _{-2.3}$ & $0.7 ^{+0.1} _{-0.2}$ \\ 
		2 & 1 & $-1.30 ^{+0.23} _{-0.16}$ & $5.4 ^{+2.3} _{-2.3}$ & $0.7 ^{+0.1} _{-0.2}$ \\ 
		\hline
		\multicolumn{5}{|c|}{field-galaxy DTD \citep{Maoz2017}} \\
		\hline
		--- & 2 & $-1.1 \pm 0.1$ & $1.3 \pm 0.1$ & $0.2 \pm 0.0$ \\ 
		\hline
	\end{tabular}
\\
\raggedright{
Notes---(1) number of redshift bins; (2) number of free parameters for the power-law fit; (3) best-fit power-law index $\alpha$; (4) Hubble-time-integrated SN~Ia number per formed stellar mass, either best fit (for the two-parameters fit) or fixed constraint (for the one-parameter fit), in $10^{-3}~{\rm M_\odot}^{-1}$; (5) DTD normalisation (equivalent to [4]), in terms of the DTD rate at a delay of 1~Gyr, i.e. $\Psi(t)=\Psi_1 (t/{\rm 1~Gyr})^\alpha$, in units of $10^{-12}~{\rm yr^{-1} M_{\odot}^{-1}}$. }
\end{table}

For comparison to other published SN rates, but also to permit future modeling of the DTD that relaxes the assumption of a single-burst SSP and allows for more extended star formation histories for the clusters, we also calculate the SN~Ia rate per unit $B$-band stellar luminosity, $R_{\rm Ia, L_B}$ (the current derivation of rates per unit formed stellar mass {\it presumes} a SSP history).  
Each cluster's rest-frame $B$-band luminosity is calculated based on its stellar flux, by interpolating or extrapolating between the F105W and F140W bands (or F160W, when F140W is not available). The SN~Ia rate is calculated as previously, except that the formed stellar mass, $M_i$, is replaced by the $B$-band luminosity of each cluster, $L_{B,i}$,
\begin{equation}
 R_{{\rm Ia}, L_B}=\frac{ N_{\rm Ia}}{\sum_i L_{B,i} \eta_i \tau_i} .
\end{equation}
Table~\ref{tab:dtd_mes} includes the SN rates per unit blue luminosity, in units of the  solar $B$-band luminosity, that we thus obtain for our cluster sample and two redshift sub-samples, as well as the literature values from previous cluster SN rate measurements. 

\section{Conclusions}
\label{Sec:conclusions}
We have analysed the imaging data from an HST high-$z$ galaxy-cluster monitoring program, in order to measure the rate of SNe~Ia in clusters at the largest-lookback times probed to date, and hence at the shortest time delays after cluster star formation. We have discovered \TotNumofSNe ~candidate SNe in these fields and measured their multi-band light curves. 
From fitting SN~Ia light-curve models to these data and examining the candidate environments, we conclude that \NumofSurveyIas ~of the transients are likely SNe~Ia in cluster galaxies within the fully time-covered area of the cluster fields. 
An additional \NumofSurveyPossibleIas ~transients are possible cluster SNe~Ia, resulting in $N_{\rm Ia} = \TotSNeRateCount \pm \TotSNeRateCountSysErr $~(systematic~error)$\pm \TotSNeRateCountStatErr $~(Poisson~error) cluster SNe~Ia found by the experiment. 
We have conducted simulations to estimate the detection efficiency, the completeness, and the photometric errors. The number of non-cluster-Ia SNe we detect is consistent with expectations from the results of previous HST field (i.e. non-cluster-targeted) SN surveys.
We have further estimated the net light from cluster galaxies, and used SPS models to convert the light to cluster stellar mass, with due attention to the systematic uncertainties in all of these quantities. 
With these inputs, we have derived the SN~Ia rate per formed unit stellar mass in galaxy clusters at the cluster redshifts, which is effectively a direct measurement of the DTD of SNe~Ia in clusters, at the delays corresponding to the time interval between star-formation and the cluster redshifts. 

Our new measurements of the SN~Ia rate and the DTD agree well with the extrapolation of previous cluster DTD measurements to higher redshifts and to earlier delay times. 
Fitting power laws to the currently measured cluster DTD values gives overall consistent values, whether one fits both the power-law index and its amplitude, or one fits only the index and imposes constraints on the DTD time integral, based on previous observations of the iron content in clusters. This agreement strengthens the case, discussed in Section~1, that compared to the SN~Ia DTD in field galaxies, the DTD in cluster environments is higher-normalised, possibly steeper, and with a significantly larger number of SNe~Ia when integrated over a Hubble time. This result, consistently emerging over the last decade, points to some physical mechanism that enhances SN~Ia numbers in cluster environments. 

As already noted, a high overall production efficiency of SNe~Ia 
could result from a stellar population with an excess of white dwarfs, relative to $\sim 1$~M$_\odot$ stars---the stars that dominate the luminosity in old populations, and thus serve to normalise the SN rate (whether by stellar luminosity or by stellar mass). In other words, the IMF of cluster stars could be "middle-heavy" with stars in the $\sim 1-8$~M$_\odot$ range, compared to a standard IMF. Alternatively, the star-formation mode in clusters can perhaps lead, by means of the initial conditions of binary formation, or by means of binary evolution, to more close binaries of the type that evolve to become SNe~Ia.
Any of these scenarios could be driven by some physical parameter relevant to star-formation, e.g. metallicity. In any event,
The large time-integrated numbers of SNe~Ia in clusters may be a hint that could help decipher the puzzles of SN~Ia progenitor and explosion physics. 
 
\section*{Acknowledgements}
We acknowledge valuable advice and discussions with Ryan Foley, Or Graur, Saurabh Jha, Steve Rodney, Keren Sharon, and the anonymous referee. This work was supported by grant 648/12 from the Israel
Science Foundation (ISF) and by grant 1829/12 of the I-CORE
program of the PBC and the ISF.


\end{document}